\documentclass[STIX1COL]{WileyNJD-v2}
\usepackage{url}
\usepackage{moreverb}
\usepackage[numbers]{natbib}
\usepackage[resetlabels,labeled]{multibib}
\newcites{P}{Appendix A: primary studies}

\newcommand\BibTeX{{\rmfamily B\kern-.05em \textsc{i\kern-.025em b}\kern-.08em
T\kern-.1667em\lower.7ex\hbox{E}\kern-.125emX}}

\articletype{Article Type}%

\received{<day> <Month>, <year>}
\revised{<day> <Month>, <year>}
\accepted{<day> <Month>, <year>}
\usepackage{float}

	\usepackage{setspace}
    \usepackage{lineno,xcolor} 	
    
    \usepackage{longtable} 
    \usepackage{tabularx} 
    \newcolumntype{L}[1]{>{\raggedright\arraybackslash}p{#1}}
    \newcolumntype{C}[1]{>{\centering\arraybackslash}p{#1}}
    \newcolumntype{R}[1]{>{\raggedleft\arraybackslash}p{#1}}
    \usepackage{relsize} 
    	
    \usepackage{rotating}
    \usepackage{multirow}
    \usepackage{makecell}

\usepackage{xcolor,colortbl}
\definecolor{Gray}{gray}{0.85}
\definecolor{LightCyan}{rgb}{0.88,1,1}   

\usepackage{float}

\begin{document}

\begin{titlepage}

  \newcommand{\HRule}{\rule{\linewidth}{0.5mm}} 

  {\center 


  \textsc{\Large This is a preprint of a paper accepted in the journal:}\\[0.5cm] 
  \textsc{\large  Software Testing, Verification and Reliability (STVR) }\\[0.5cm] 


  \HRule \\[0.4cm]
  { \huge \bfseries Improving Test Automation Maturity: a Multivocal Literature Review}\\[0.4cm] 
  \HRule \\[0.5cm]



Early view citation:
Wang, Y, Mäntylä, MV, Liu, Z, Markkula, J, Raulamo-jurvanen, P. Improving test automation maturity: A multivocal literature review. Softw Test Verif Reliab. 2022;e1804. https://doi.org/10.1002/stvr.1804
  
    \vspace{.5 cm}
 }  







\vfill 

\end{titlepage}


\newpage

\title{Improving Test Automation Maturity: a Multivocal Literature Review \thanks{}}

\author[1]{Yuqing Wang*}

\author[1]{\large Mika V. M{\"a}ntyl{\"a}}

\author[1]{Zihao Liu}

\author[1]{Jouni Markkula}

\author[1]{P{\"a}ivi Raulamo-jurvanen}

\authormark{Yuqing Wang \textsc{et al}}

\address[1]{M3S research unit, University of Oulu, Pentti Kaiteran katu 1, Oulu 90014, Finland}



\corres{*Yuqing Wang, Corresponding address. \email{yuqing.wang@oulu.fi}}

\abstract[Abstract]{Mature test automation is key for achieving software quality at speed. In this paper, we present a multivocal literature review with the objective to survey and synthesize the guidelines given in the literature for improving test automation maturity.  We selected and reviewed 81 primary studies, consisting of 26 academic literature and 55 grey literature sources.  From primary studies, we extracted 26 test automation best practices (e.g., Define an effective test automation strategy, Set up good test environments, Develop high-quality test scripts) and collected many pieces of advice (e.g., in forms of implementation/improvement approaches, technical techniques, concepts, experience-based heuristics) on how to conduct these best practices. We made main observations: (1) There are only 6 best practices whose positive effect on maturity improvement have been evaluated by academic studies using formal empirical methods;  (2) Several technical related best practices in this MLR were not presented in test maturity models; (3) Some best practices can be linked to success factors and maturity impediments proposed by other scholars; (4) Most pieces of advice on how to conduct proposed best practices were identified from experience studies and their effectiveness need to be further evaluated with cross-site empirical evidence using formal empirical methods; (5) In the literature, some advice on how to conduct certain best practices are conflicting, and some advice on how to conduct certain best practices still need further qualitative analysis.}


\maketitle


  
\section{Introduction}\label{sec:introduction}
\noindent  
Modern software development approaches, such as Agile, DevOps, and continuous integration, have changed the way software is being developed \cite{worldqualityreport19,pressman2005software}. Software products are released faster and more frequently than before - software release frequency improved from around 12 months in the last century to 3 weeks these days \cite{pressman2005software}. How to ensure development speed without sacrificing quality is essential. Quality issues may lead to disappointments of customers, reputation damage of software products, and security problems - thus, the loss of the market \cite{bubevski2014novel}. Test automation is key for achieving `software quality at speed' in modern software developments, because of the rapid speed it offers to test software products in short test cycles \cite{Mika,worldqualityreport19}. With the rise of modern software developments, test automation is growing in popularity. According to the statistics~\cite{market19}, ``the global test automation market size is expected to grow from USD 12.6 billion in 2019 to USD 28.8 billion by 2024''.

However, many organizations still have immature test automation practices with negative outcomes, e.g., unable to automate valuable tests, create and maintain automated tests at optimal costs, or detect defects on time. Such negative outcomes impede organizations to reap the expected test automation benefits, consume software development resources, and even risk software quality.  The software industry and research community refer to "improving immature test automation practices with negative outcomes" as "improving test automation maturity" \cite{ISTQB_Glossary,icsoft20,garousi2016Decison}. The software industry has seen increasing attempts to improve test automation maturity but not all attempts are effective \cite{ISTQB2020,ISTQB2018}. Based on ISTQB's recent software testing practice survey \cite{ISTQB2020}, around 65\% of the almost 2000 software organizations in the world attempted to improve test automation maturity while only about a half reported that their attempts are effective. The industry needs the guidelines to drive for effective test automation maturity improvement \cite{wiklund2015impediments,ISTQB2020,ISTQB2018}

 The current literature includes many guidelines for improving test automation maturity~\cite{Mika,garousi2017taNot,wiklund2017impediments}. Surveying and synthesizing the guidelines given in the literature is essential to integrate and link the guidelines from different sources into a whole. Yet, the research effort on that is limited \cite{Mika,Wang2019}. Our previous work~\cite{Wang2019} stands for the purpose to narrow this gap. Our previous work is the first attempt to survey the existing guidelines and synthesize test automation best practices. It reviewed 18 test maturity models that are used to guide test automation practices in the industry. From these 18 test maturity models, our previous work extracted some test automation best practices but failed to identify qualitative analysis on these best practices and how they can be conducted. This calls for a systematic literature review (SLR) on a larger pool of sources to synthesize test automation best practices with further qualitative analysis.

This paper presents a multivocal literature review (MLR) with the objective to survey and synthesize the guidelines given in the  literature for improving test automation maturity. It intends to complement the current  research that lacks a SLR of a large pool of sources to synthesize test automation best practices with further qualitative analysis. MLR is a type of SLR that includes both academic literature (AL) and grey literature (GL). Our reason to use the MLR approach is the industrial relevance of our topic. Based on prior scholars \cite{ricca2021web,garousi2018guidelines}, GL (e.g.,  white papers, magazines, blog posts, wikis, and technical reports) is ``still an unexplored gold mine of guidelines for test automation'', as practitioners exist in large numbers and they are sharing their experience matured on the field and propose test automation guidelines. Surveying the guidelines for improving test automation maturity from both AL and GL is critical to enable knowledge transfer between industry and research and identify the gaps.

In this MLR, we searched the AL using Google Scholar, Scopus, and Web of Science, and the GL using the Google search engine on the topic of test automation maturity improvement. From a large pool of sources, we selected 81 primary studies (consist of 26 AL and 55 GL sources) and studied them. As the main contributions, from these 81 primary studies, we extracted 26 test automation best practices using the thematic analysis approach, and collected many pieces of advice (in forms of implementation/improvement approaches, actions, technical techniques, concepts, experience-based heuristics) on how to conduct these best practices. Using cited sources, we explained these 26 best practices and the advice around them. We have several observations related to these 26 best practices and the advice around them:

\begin{itemize}
\setlength\itemsep{-0.3em}
   \item  There are only six best practices (Define an effective test automation strategy, Provide enough resources, Have competent test professionals, Select the right test tools, Set up good test environments, Design the SUT for automated testability) whose positive effect on maturity improvement have been evaluated by academic studies using  formal empirical methods.

    \item This MLR identified six technical related best practices (Set up good test environment, Create high-quality test data, Develop high-quality test scripts, Automate test oracles, Analyze test automation results efficiently and effectively, Adopt new technologies), which were not presented in test maturity models reviewed in our previous work \cite{Wang2019}.
    
    \item Some best practices identified in this MLR can be linked to test automation success factors and maturity impediments proposed by other scholars. 
    
    \item  Most pieces of advice on how to conduct proposed best practices were identified from experience studies and their effectiveness need to be further evaluated with cross-site empirical evidence using formal empirical methods. 
    
    \item In the literature, some advice on how to conduct several best practices are conflicting, e.g., as the advice for "Select the right test tools", many scholars suggested selecting test tools against pre-defined selection criteria, while some practitioners viewed that - based on their experience - selecting test tools against pre-defined criteria is less useful than selecting each test tool with an experimentation mindset.
    
    \item In the literature, some advice on how to conduct certain best practices still need further qualitative analysis, e.g.,  as the advice for "Use the right test automation metrics", the literature defined the concept of "right test automation metrics" and illustrated a collection of example ones, but it did not mention how to define and customize test automation metrics based on own needs of organizations.
\end{itemize}

Our study provides a single source that surveyed and synthesized the guidelines given in both AL and GL for test automation maturity improvement. It can be useful to both practitioners and researchers, who can, respectively, consult our 26 test automation best practices with the advice on conducting them for improving test automation maturity, and foster future research in this field. 

The remainder of this paper is structured as follows. Section~\ref{sec:background} introduces the concepts, explained MLR approach, and reviews related work. Section~\ref{sec:research method} describes the research process and methodology used to conduct this MLR. Section~\ref{sec:results} reports the study results. Section~\ref{sec:dicussion_sec} summarizes and discusses the study findings, explores the implications of this MLR to research and practice, and examines threats to validity. Section~\ref{sec:conclusion} concludes the paper and states its contributions. 

\section{Background}\label{sec:background}
\noindent In the following subsections, we study the general concept of maturity and software test automation maturity in Section~\ref{sec:conceptOfMaturity}, introduce the MLR approach in software engineering (SE) in Section \ref{sec:background_MLR}, and review related work in Section~\ref{sec:relatedwork}. 

\subsection{The concept of maturity and software test automation maturity}
\label{sec:conceptOfMaturity}
According to the Oxford dictionary \cite{stevenson2010oxford}, maturity is defined as: ``the state of being fully grown or developed". The concept of maturity has been introduced into various fields. For instance, software process maturity in SE \cite{paulk1995}, business process maturity in the area of business management~\citep{tarhan2016}, big data maturity in data science. In a literature review study~\cite{mettler2011}, Mettler has studied the concept of maturity in different fields and concluded three forms of maturity:

\begin{itemize}
\setlength\itemsep{-0.3em}
    \item Process maturity,  a specific process ``is explicitly defined, managed, measured, controlled, and effective".
    
    \item Object maturity,  a particular object ``like a software product, a machine or similar reaches a predefined level of sophistication".
    
    \item People capability, the work force ``is able to enable knowledge creation and enhance proficiency".
\end{itemize}


Based on test automation literature and Mettler's views of maturity, test automation maturity consists of three dimensions; each dimension is dependent on the other. Thus, this MLR considers test automation maturity improvement with those three dimensions as the whole: 

\begin{itemize}
\setlength\itemsep{-0.2em}
    \item Test automation process maturity: \textit{a test automation process is explicitly defined, managed, measured, controlled, and effective} \cite{garousi2017taNot,pocatilu2002,icsoft20}. A test automation process is a sequence of test automation activities \cite{garousi2017taNot,pocatilu2002}. Test automation process maturity demonstrated that all test automation activities are organized in a structured process to produce expected outcomes~\cite{garousi2017taNot,pocatilu2002,softwareTestAutomation}.
    
    \item Test automation technology maturity: \textit{test automation technology reaches a predefined level of sophistication to meet the needs of software development} \cite{ammann2016,wiklund2017impediments,softwareTestAutomation}. Test automation technology is a set of techniques, expertise, tools, methods, and processes that can be used in the development and application of test automation \cite{softwareTestAutomation}. 
   Test automation is technology-driven \cite{wiklund2017impediments}. The increased level of sophistication of test automation technology  allows the increased test scale, test coverage, test efficiency, and test effectiveness \cite{ammann2016}.
   
   \item People capability: \textit{the workforce is able to leverage the competencies in test automation} \cite{garousi2020,Mika,tmap}. A variety of expertise is required for people to be able to work on test automation \cite{Mika,tmap}. People capability concentrates on continuously improving the management and development of human assets to ensure that the right people with enough expertise are playing the roles in test automation \cite{garousi2020,tmap}. 
\end{itemize}

\subsection{Multivocal literature review}
\label{sec:background_MLR}
SLR is a popular research methodology to conduct literature reviews for aggregating all existing evidence on research questions SE~\cite{kitchenham2009}. MLR is a type of SLR that includes both AL and GL sources (e.g. white papers, web pages, magazines, technical reports, wikis, and blog posts)~\cite{garousi2018guidelines}. Recently, MLR has been actively applied in SE research and is attracting attention \cite{garousi2016need}. Prior MLRs have made contributions to SE research. For instance, MLR  \cite{Sgarousi2016} contributed five groups of factors to support automation decisions in software testing; MLR \cite{garousi02testmaturitymodel} identified 58 test improvement models and contributed the drivers, challenges, and benefits of using these models; MLR \cite{myrbakken2017} contributed the concept, characteristics, benefits, challenges of DevSecOps.

As concluded by Garousi et al.~\cite{garousi2018guidelines}, SE research needs MLRs for many reasons. First, GL is the source of up-to-date research, as SE practitioners are actively producing fresh GL sources on a large scale. Second, neglecting GL may not able to provide enough insights into ``the state of the art and practices” in the current SE context. Third, SE research is practical-oriented, thus, it is essential to hear the voice of both SE researchers and practitioners in order to narrow the gap between academia and industry. In this paper, our reason to use the MLR approach relates to the industrial relevance of our topic and is consistent with prior SE researchers who conduct MLRs.

\subsection{Related work}
\label{sec:relatedwork}

Some secondary studies have surveyed the literature on test automation maturity improvement. A summary of these studies is in Table \ref{tab:relatedSecondary}. In Table \ref{tab:relatedSecondary}, the main contributions of each study are presented in column ``Contributions''. By further investigation, we found that the contributions of each of these studies are insufficient to guide the industry to improve test automation maturity, the specific reasons for each study are listed in column ``Gaps''. How our MLR narrows the gap of the prior one is concluded in column "Our MLR".

{
\begin{longtable} {p{0.4cm} p{1.1cm} p{1.86cm} p{4.2cm} p{3.5cm} p{4cm}}
  \caption{ \label{tab:relatedSecondary} Relevant secondary studies}
    \\ \hline
   Ref. & Method &  Primary study & Contributions & Gaps & Our MLR \\ \hline
   
   \cite{Mika2012} 

   & SLR and Survey & 25 AL sources & This study found that maximizing the benefits (improve product quality, test coverage, reliability, fault detection) and addressing drawbacks (lack of skilled people, test automation needs time to be mature) can improve test automation maturity. & A taxonomy of benefits and drawbacks is too high level. How to maximize proposed benefits and address proposed drawbacks in  practice was not mentioned. & Our MLR proposed 26 test automation best practices (grouped into 13 key areas) that are more practical. It also collected the advice of prior researchers and practitioners about how to conduct each best practice. 
    \\\hline
   
    \cite{wiklund2017impediments} 
     & SLR & 39 AL sources & Five categories of impediments that hinder test automation maturity: behavioral effects, business and planning, skills, test systems, system under test. & Five categories of impediments are too high level and miss some test automation key areas (Test design, Test execution, Verdicts, and Measurements). How to tackle those implements for improvements were barely mentioned. &  Our MLR proposed 26 test automation best practices (grouped into 13 key areas) that are more practical. Our MLR concerned test automation key areas (Test design, Test execution, Verdicts, and Measurements) that were not covered in their study. Best practices in our MLR can be used to tackle the relevant impediments in their work.  \\ \hline
    
    \cite{rodrigues2016} 
   & LR and Survey & 4 AL sources & 12 critical success factors of test automation: feasibility assessment, testability level of the software under test, resource availability, manageability, well defined test process, scalability, maintainability, automation tool acquisition criteria, quality control, resource reusability, dedicated and skilled team, automation planning/strategy. & Not a SLR. Their study missed some test automation key areas (Test environment, Test automation requirements, Test design, Test execution, Verdicts). How to improve test automation maturity against those critical success factors was not introduced. & Our MLR is a SLR. Our MLR concerned test  automation  key areas (Test environment, Test automation requirements, Test design, Test execution, Verdicts) that were not covered in their study. Best practices in our work can be used to achieve relevant success factors in their work.     \\\hline
     
   \cite{Mika} 
     & SLR & 26 AL + 52 GL sources & 5 types of factors affect when/what to automate in software testing decisions: software under test-related factors, test-related factors, test-tool-related factors, human and organizational factors, cross-cutting and other factors.
    & This study only addresses what and when to automate decisions; It did not advise the improvement steps after initial decisions in a test automation process. & Our MLR is a SLR. Our work advises best practices throughout the whole lifecycle of a test automation process.\\ \hline   
   
    \cite{Wang2019}   & LR & 18 test improvement models and expert reviews & Test automation best practices in 13 key areas: test automation strategy, resources, test organization, knowledge transfer, test tools, test environment, test automation requirements, test design, test execution, verdicts, test automation process, measurements, SUT.
    &  Not a SLR. It only named test automation best practices. Why each best practice is suggested and how it can be conducted were not included.  & Our MLR proposed test automation best practices with a SLR. We identified several technical related best practices that did not present in their work. We explained the proposed best practices and collected many pieces of advice on how to conduct proposed best practices.     \\\hline   
\end{longtable}}

To sum up, our MLR complements prior studies and makes the novelty from the following aspects: 
\begin{itemize}
\setlength\itemsep{0em}
    \item This MLR concerned 13 test automation key areas to study test automation maturity improvement. Compared to most previous secondary studies, this MLR concerned more test automation key areas.
    
    \item Our MLR reviewed both AL and GL sources on test automation maturity improvement. Due to the industrial relevance of this research topic, including the GL is necessary. Only one prior secondary study \cite{Mika} included GL and that study was about what and when to automate in software testing.
    
    \item This MLR proposed test automation best practices that allow for improving test automation maturity. Our previous work~\cite{Wang2019} is the only one that also proposed test automation best practices. This MLR identified several technical related best practices that were not presented in our previous work, see details in Section \ref{sec:SumaaryStudyResults}. This MLR complements our previous work by exploring why each best practice is suggested and collecting the advice of prior researchers and practitioners on how to conduct proposed best practices.
    
    \item This MLR was recently finished and intended to provide the current view for this research scope. Prior secondary studies were published in 2016 or earlier years.
\end{itemize}

\section{Research method}
\label{sec:research method}
 Our MLR was carried out following `the guidelines for conducting multivocal literature reviews' from Garousi et al.~\cite{garousi2018guidelines}.  Our research process contains six stages: Review question, Search strategy, Source selection process, Quality assessment, Mapping of primary studies, and Data extraction and synthesis. Each stage is described below.

\newcommand{\RQoneTitle}{RQ1 -- Key areas}
\newcommand{\RQone}{What key areas test automation activities are clustered in?}

\newcommand{\RQtwoTitle}{RQ2 -- Practices}
\newcommand{\RQtwo}{What practices should be adopted to mature test automation?}

\newcommand{\RQthreeTitle}{RQ3 -- Maturity models}
\newcommand{\RQthree}{Which test automation maturity models have been developed?}

\newcommand{\RQfourTitle}{RQ4 -- Trends}
\newcommand{\RQfour}{What are the trends of this research area?}

\newcommand{\RQfiveTitle}{RQ5 -- Research \& contribution types}
\newcommand{\RQfive}{ What types of studies have been carried out in this area to make contributions?} 

\subsection{Review question}
\label{sec:reviewQuqestions}
To address the objective of this MLR, we defined two review questions:

\begin{itemize} 
\item \textbf{RQ1. Which test automation best practices are given in the literature?} This review question aims to extract a taxonomy of test automation best practices from the literature. 

\item \textbf{RQ2. What advice was given in the literature about how to conduct proposed test automation best practices?} This review question attempts to explore the advice in the literature about how to conduct proposed test automation best practices in RQ1.
\end{itemize} 

\subsection{Search strategy}
 The search strategy was developed to search for relevant AL and GL sources to answer our review questions in this MLR. The first author developed the search strategy and it was reviewed and revised by other co-authors. The search strategy defined the search string, search databases, inclusion and exclusion criteria, and the source selection process.

\paragraph{Search string.} Our search string was defined by following the standard steps proposed by Brereton et al.~\cite{brereton2007lessons}. Five major terms were defined according to our research topic: software, test, automation, maturity, improvement. Alternative synonyms of those major terms were identified by reviewing prior literature. The search string was then defined accordingly, see Table~\ref{tab:searchStrings}. The complete search string is formulated by combining the rows with AND. 

\begin{table}[h]
\centering
\caption{Search string}\label{tab:searchStrings}
\begin{tabular}{l l l l}
\hline
 Row & Terms \\
	\hline
	A & software \\ \hline
    B & test OR testing \\ \hline
    C & automation OR automated  \\ \hline
    D & mature OR maturity OR maturation OR matureness \\ \hline
    E & improvement OR improve OR assessment OR assess \\ 
\hline
\end{tabular}
\end{table}


\paragraph{Search databases.} We conducted a preliminary search in different bibliographic databases and search engines. Search databases were selected depending on the quality and availability of relevant sources on them. We decided to use Scopus, Web of Science, and Google Scholar to search for AL sources. With our search string, Scopus and Web of Science can return many quality AL sources relating to our research topic. Google Scholar allows retrieving the full-text of AL sources from open access journals and pre-print repositories. The Google search engine was chosen to search for GL sources. It can return more relevant GL sources with our search string compared to other search engines. However, prior researchers have argued that using search engines may introduce the selection bias into SLRs, because the ranking algorithm of search engines may tailor search results based on user preferences and behaviors~\cite{curkovic2018}. Studies \cite{wang2016,mowshowitz2005} have introduced the approaches to reduce such selection bias in SLRs. In this MLR, we took the following approaches:   

\begin{itemize}
    \item  The search in Google Scholar or Google search engine was done without logging into a Google account. The search history and cache in the browser were cleared up before the search.
    
    \item After applying the search string in Google Scholar or Google search engine, search results were exported into a spreadsheet that stores the title and URL of each source. In the source selection process, we open each URL to screen the sources. 
\end{itemize}


\paragraph{Inclusion and exclusion criteria.}
\label{sec:criteria}
Table~\ref{tab:selectionCriteria} shows inclusion criteria to select relevant sources for this MLR. Exclusion criteria are the reverse of inclusion criteria. This means a source that does not meet inclusion criteria will be excluded.

\begin{table}[h]
\centering
\caption{Inclusion criteria}\label{tab:selectionCriteria}
\begin{tabular}{ p{13cm} }
\hline
	C1.The AL or GL source is written in English. \\
	
	C2.The AL or GL source is full-text accessible.\\
	
	C3.The AL or GL source is not a duplication of others.\\
	
	C4.The AL source was published in journals, conferences, workshops, and books. \\
	
	C5.The GL source was published in blogs, magazines, web pages, white papers, technical reports, tutorials, wikis, presentations, broadcastings, videos. \\
	
	C6.The AL or GL source is relevant to test automation maturity improvement \\
\hline
\end{tabular}
\end{table}


\subsection{Source selection process}
\begin{figure}[ht]
  \centering
  \includegraphics[width=0.9\linewidth]{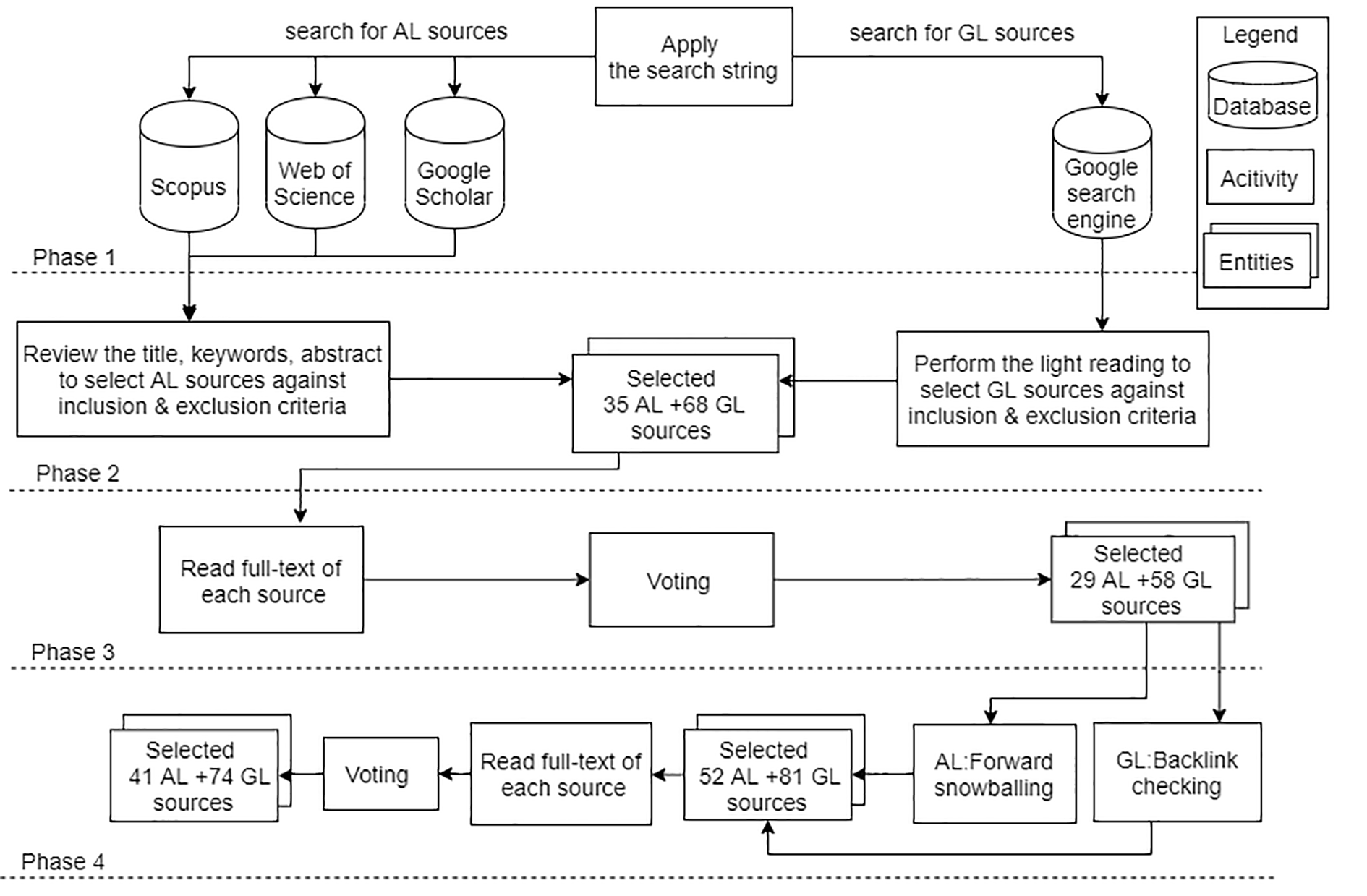}
  \caption{An overview of search selection process.}
  \label{Fig_searchProcess}
\end{figure}

The search selection process consists of four phases, see Figure~\ref{Fig_searchProcess}. Table~\ref{tab:searchResults} summarizes the search results of different phases. In phase 1, our search string was applied in the chosen search databases. It was modified to fit the search format of each database. The search results were shown in each database.

\begin{table}[h]
\centering
\caption{Summary of search results}\label{tab:searchResults}
\begin{tabular}{ lllll }
\hline
databases & Phase 1 & Phase 2 & Phase 3 & Phase 4 \\
\hline
	Scopus & 76 & 5 & 2 & 3\\ \hline
	Web of Science & 32 & 2 & 1 & 1\\ \hline
	Google Scholar & 354,000+ & 28 & 26 & 37\\ \hline
	Google search engine & 32,500,000+ & 68 & 58 & 74 \\ 
\hline
\end{tabular}
\end{table}

In phase 2, sources appearing in search results of phase 1 were reviewed against the defined inclusion and exclusion criteria (section~\ref{sec:criteria}).  The title, abstract, and keywords were reviewed to screen AL sources. Light reading was performed to screen GL sources. Note that, though there were more than 354000 search results in Google Scholar and 32500000 search results in Google search engine, the relevant sources only presented in the first few pages. Thereby, in Google Scholar and Google search engine, sources on  first-ten pages were screened. At the end of this phase, 35 AL and 68 GL sources were selected for further review.

In phase 3, the first and third authors read  full-text of each source selected from phase 2. They used `yes' or `no' to vote for whether a source meets each inclusion criterion. Only the source got `yes' for all inclusion criteria from each author were selected. In the end, 29 AL and 58 GL sources were selected to build up our initial pool of sources.

In phase 4, we used Forward snowballing approach to screen additional AL sources based on the citations of selected AL sources in phase 3, and checked the backlinks of selected GL sources in phase 3 to screen additional GL sources.  This screened additional 23 AL and 23 GL sources.  The first and third authors read full-text of each of these sources against the defined inclusion and exclusion criteria~(Section~\ref{sec:criteria}). They used `yes' or `no' to vote for whether a source meets each inclusion criterion. Only 12 AL and 16 sources got `yes' for all inclusion criteria from each author were selected. Our final pool consists of 41 AL and 74 GL sources.

\subsection{Quality assessment}
The quality of sources in our final pool was assessed to minimize bias and maximize the validity of this MLR. Since AL and GL sources do not follow the same review and publication process \cite{yasin2020}, a separate checklist was created to assess the quality of each type of sources.

To assess the quality of AL sources, we studied quality assessment guidelines from Zhou et al.~\cite{zhou2015quality} and Kitchenham~\cite{kitchenham2004procedures}. We developed a checklist that contains 20 quality-check questions for assessing the quality of AL sources in this MLR. Besides, Garousi et al.'s GL quality assessment checklist~\cite{garousi2018guidelines} was used to assess the quality of GL sources in this MLR. We revised their checklist to fit our needs. Our checklist for GL sources contains 17 quality-check questions. Our checklists for both AL and GL sources are presented in: https://doi.org/10.6084/m9.figshare.13554164.v1

To rate the quality of each source,  we used 3-point-scale (1= yes, 0.5 = partly, 0 = no) to answer quality-check questions one by one in each of our checklists. This method has been used in prior studies, e.g.,~\cite{Mika,garousi02testmaturitymodel}. The total score of each source was calculated. Sources scored 50\% or more of the maximum points were finally included.  This means, to be included, the AL source should get at least 10 out of 20 points, while the GL source should get 8.5 or more out of 17 points. As a result of quality assessments, 15 AL and 19 GL sources were excluded. The rest 81 sources (including 26 AL and 55 GL sources) were used as primary studies for this MLR. For the sake of transparency, we listed out all 81 primary studies in Appendix A: P1-P26 are AL sources while P27-P81 are GL sources. Many highly cited SLR guidelines (e.g., \cite{petersen2015guidelines,garousi2018guidelines,keele2007guidelines}) recommended listing out primary studies in the appendix - that could let readers easily distinguish primary studies from other cited sources in a research paper. Many previous SLRs (e.g.\cite{Mika,wiklund2017impediments,Sgarousi2017}) have followed such guidelines to do so. Besides, our paper repository can be found in: https://doi.org/10.6084/m9.figshare.17059925.v1.

\phantom{\citeP{Srodrigues2016relevance,Sgarousi2020exploring,Skoomen2013tmap,furtado,Spersson2004establishment,Skoomen1999test,SsoftwareTestAutomation,Skhan2018issues,Sgarousi2016,Sgarousi2017,Swiklund2017impediments,Swang2020survey,Sricca2021web,Skasurinen2010software,Slee2012survey,Sdeak2013organization,Swang2020test,Sborjesson2012,raulamo2019,Skasoju2013,Silles,Shaugset2011,Swiklund2012,SSri2014,garousi2017test,SMika2012}}

\subsection{Mapping of primary studies in our final pool}
\label{sec:mappingstudyresults}
We mapped our 81 primary studies in four aspects: Number of sources per year, Numbers of sources by source type, Author profile, and Number of sources by contribution and research type.

\begin{figure}[hpb]
\centering
  \includegraphics[width=0.76\linewidth]{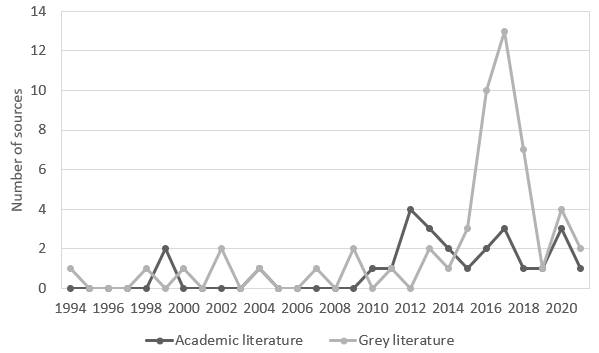}
  \caption{Number of sources per year}
  \label{fig:figure2}
\end{figure}

 \paragraph{Number of sources per year.} Figure~\ref{fig:figure2} shows the number of sources per year by literature type. The identified sources were produced between 1994 and 2021. Before the year 2012, there were less than 2 sources per year for both GL and AL sources. More  sources were identified in a recent decade.

\begin{figure}
\centering
  \includegraphics[width=0.7\linewidth]{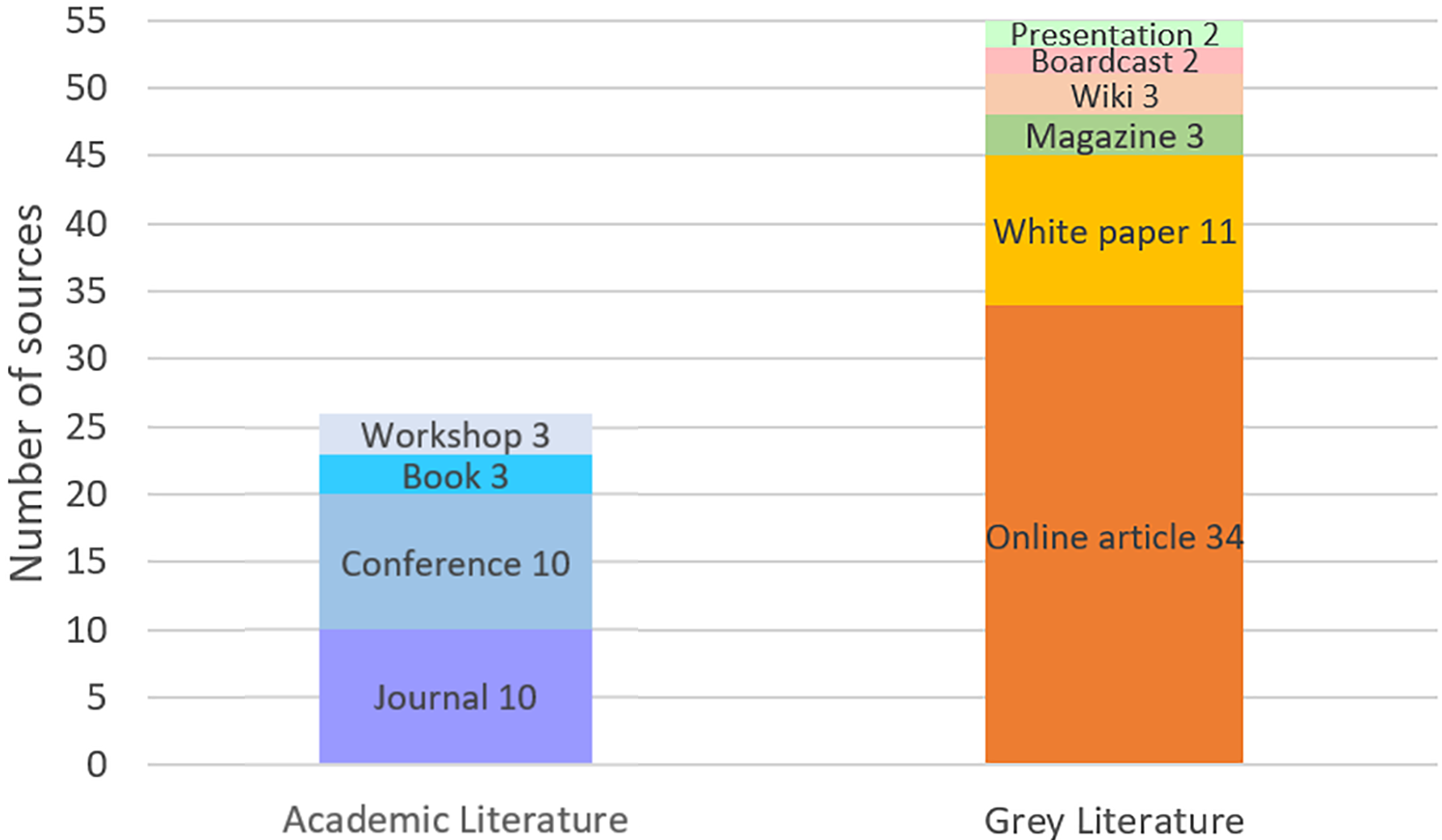}
  \caption{Number of sources by type.}
  \label{fig:sourceType}
\end{figure}

\paragraph{Numbers of sources by source type.} Figure~\ref{fig:sourceType} shows the number of sources by source type in both categories: AL versus GL. As noted in this figure, conference papers (n=10) and journal papers (n=10) take a large proportion of the AL sources. Most GL sources are  online articles (n=34) published in blogs, technical websites, and magazines.

 \paragraph{Author profile.} All 26 AL sources have at least two authors. 12 AL sources have the author, who specializes in software testing and has more than a hundred publications. Most of those senior researchers move their interest to software test automation in recent years. Table \ref{tab:authorProfile_GL} depicts the author profile of 55 GL sources in this MLR. As shown in this table, the majority of GL sources (n=39) were produced by identifiable individual practitioners.

\begin{table}[htpb]
\centering
\caption{Author profile of grey literature sources.}\label{tab:authorProfile_GL}
\begin{tabular}{ ll }
\hline
    & N \\ \hline
    \textbf{Identifiable individual practitioner:} & \\
    Test Lead/Manager/Director & 11 \\
    Tester & 6\\
    Developer & 2 \\
    Consultant & 15 \\
    Others & 5 \\
    \textbf{Identifiable organization} & 13 \\
    \textbf{Unsigned author} & 3\\
    
\hline

\hline
\end{tabular}
\end{table}

 \paragraph{Number of sources by contribution and research type.} To analyze contribution types of our primary studies, we used the categories from Garousi et al.~\cite{garousi2018guidelines}: Heuristics \& Guidelines, Model, Process, Method, Metric, and Tool. To analyze the research type of our primary studies, we studied the guidelines from Garousi et al.~\cite{garousi2018guidelines} and Petersen et al.~\cite{petersen2008systematic},
and accordingly classified our primary studies into four categories:
 
\begin{itemize}
\setlength\itemsep{-0.3em}
    \item Evaluation (Eva): These studies use formal empirical methods (e.g., practitioner survey, case studies, controlled experiments, hypothesis testing) to evaluate test automation maturity improvement related findings. 
    \item Experience (Exp): These studies present test automation maturity improvement related findings based upon the authors' (can be individuals or software organizations) experience and expertise. 
    \item Opinion (Opi): These studies present test automation maturity improvement related findings based on authors' (can be individuals or software organizations) opinions. 
    \item Other: These studies do not fit into any of the above research types, including philosophical studies and survey studies. 
\end{itemize} 

\begin{figure}[h!] 
\centering
  \includegraphics[width=0.7\linewidth]{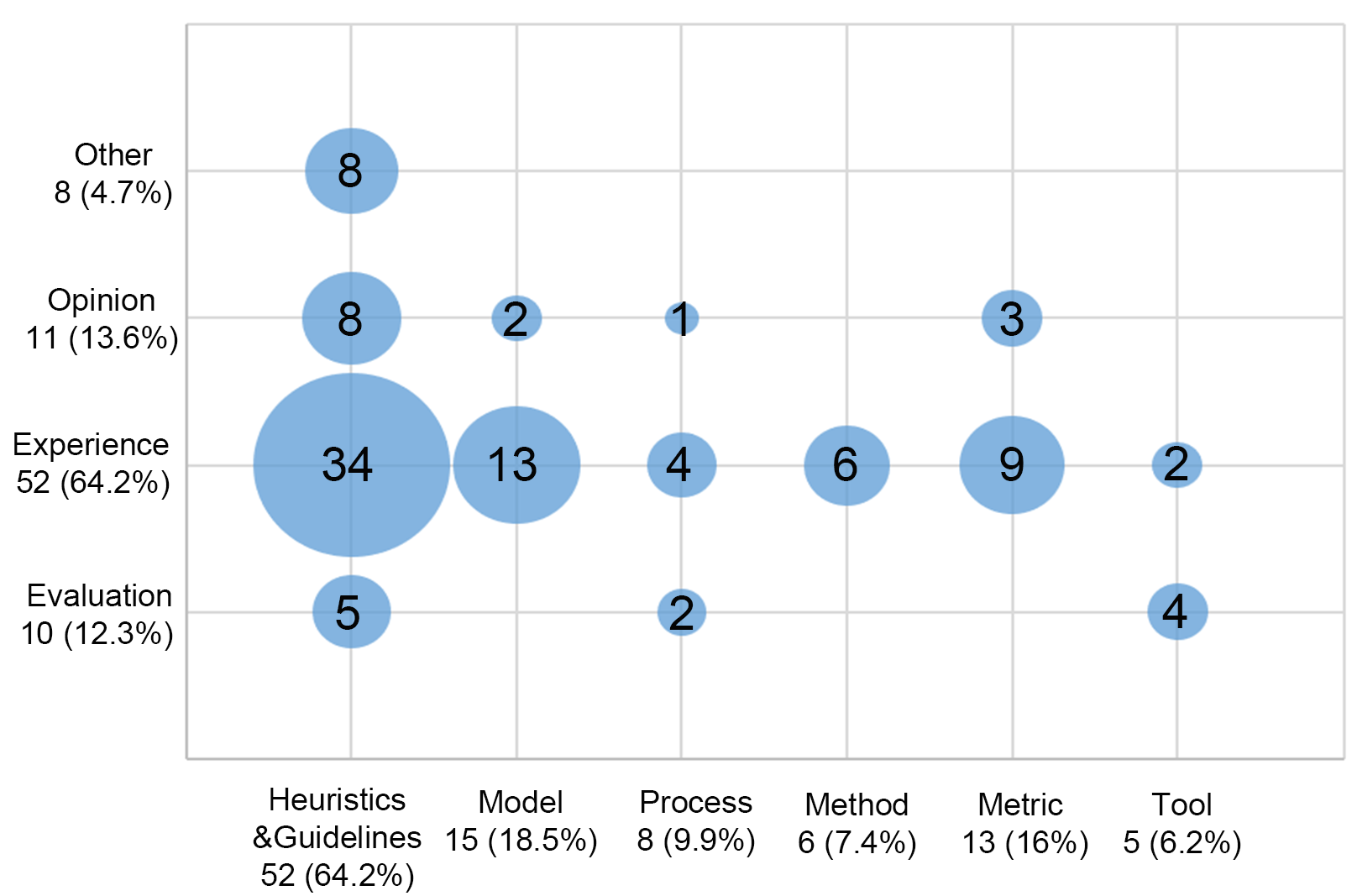}
  \caption{Mapping primary studies by contribution and research type}
 \label{fig:bubbleplot}
\end{figure}

 Figure~\ref{fig:bubbleplot} shows the mapping of our primary studies by contribution and research type. Note that, since a source can have multiple types of contributions, the total number of studies (n=99) in all contribution types is greater than the total number of primary studies (n=81). Only 10 primary studies are evaluation studies, while 4 of them purely focus on test tool selection and usage related topics. Many of our primary studies are experience studies (n=34) that contributed to test automation heuristics \& guidelines. 15 out of our primary studies proposed test maturity models. 4 test maturity models were proposed by authorized organizations: TMap (proposed by Sogeti), TPI (proposed by Sogeti), STBox~3.0 (proposed by CTG Europe), TestSPICE~3.0 (proposed by Intacs). The rest models were proposed by individual practitioners or software organizations based on experience and expertise.

\subsection{ Data extraction and synthesis}
\label{sec:dataExtracton}
We coded our primary studies using an integrated deductive and inductive approach proposed by Cruzes and Dyba \cite{cruzes2011}. A data analysis software NVivo\footnote{NVivo: https://www.qsrinternational.com/nvivo-qualitative-data-analysis-software/} was used to assist the coding process.  

\begin{figure}[h!]
\centering
  \includegraphics[width=0.45\linewidth]{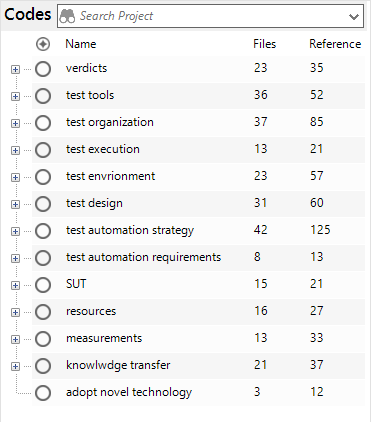}
  \caption{NVivo screenshot 1: coding status at the initial phase}
  \label{fig:NVivo1}
\end{figure}

We read the full content of each primary study to identify the findings that propose test automation best practices. Relevant findings were coded from each primary study. Our prior work \cite{Wang2019} reviewed 18 test maturity models and identified 13 test automation key areas. In this MLR, pre-defined code categories (to classify test automation best practices) were created according to these 13 test automation key areas identified in our prior work: test automation strategy, resources, test organization, knowledge transfer, test tools, test environment, test automation requirements, test design, test execution, verdicts, test automation process, measurements, software under test (SUT). Pre-defined code categories were allowed to be modified depending on actual coding situations. All codes at this stage were assigned into at least one code category. Figure \ref{fig:NVivo1} is an NVivo screenshot showing coding status at this stage. In this screenshot, ``Name'' represents code categories; ``Files'' counts the number of cited sources; ``References'' counts the total number of codes on all cited sources. The total number of files is less than the sum of references, as a single source may have many codes.

\begin{figure}[ht]
\centering
  \includegraphics[width=1\linewidth]{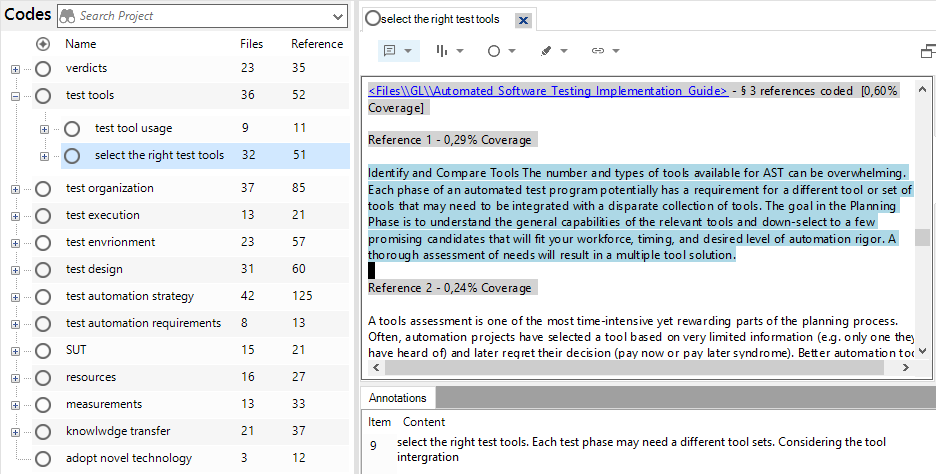}
  \caption{NVivo screenshot 2: examples of high level code categories, codes, and annotations}
  \label{fig:Nvivo2}
\end{figure}

 All codes were reviewed by reading through the content they included. The content of each code was displayed and examined in the original context of a cited source, considering the surrounding texts of the code. An annotation was written to conclude a concise description of each code. Based on codes and annotations, we created high-level categories upon initial code categories using an inductive approach.  Each high-level code category represented a test automation best practice. All codes were assigned into at least one high-level category. Figure \ref{fig:Nvivo2} shows an NVivo screenshot that gives examples: high-level categories "select the right test tools" and "test tool usage" were created under the initial code category "test tools"; Example codes were assigned to the high-level category "select the right test tools"; An annotation concluded a code.

Based on codes in high-level coding categories, a taxonomy of test automation best practices was formulated. We refined best practices based upon the content of codes, annotations, and cited sources until a final taxonomy was obtained. By reviewing the content of codes, annotations, and cited sources again, we collected the advice on how to conduct these best practices. To answer our RQ1, we intended to present our final taxonomy of test automation best practices and demonstrate cited sources by source type (AL and GL) and research type (Eva, Exp, Opi, Other). To answer our RQ2, using cited sources, we intended to explain each best practice and present the advice on how to conduct it. The explanation and advice to best practices were distinguished by source type (AL and GL) and research type (Eva, Exp, Opi, Other).

\section{Results}
\label{sec:results}
In the following sub-sections, we present the study results to our review questions in turn.
\subsection{RQ1 - Test automation best practices}
\label{sec:resultsPractices}

We extracted 26 test automation best practices from our 81 primary studies (that consist of 26 AL and 55 GL sources). Table~\ref{tab:factors} shows these 26 best practices grouped into 13 key areas. We had two observations on Table~\ref{tab:factors}. First, we found that, some best practices (``Adjust the strategy to the changes'', ``Automate test oracles'', ``Adopt new technologies'') suggested in the GL did not be suggested in the AL from our primary studies. Second, we noticed that, there are only 6 best practices (``Define an effective test automaton strategy",``Provide enough resources'', ``Have competent test professionals'', ``Select the right test tools'', ``Set up good test environments'', ``Design the SUT for automated testability'') whose positive effect on test automation maturity have been validated by academic evaluation studies (from our primary studies) using formal empirical methods.

{
\begin{longtable}{|p{2.33cm}|p{3.7cm}|p{4cm}|p{5cm}|c|  } 
\caption{\label{tab:factors} Best practices of test automation} \\ \hline
     \textbf{Key areas} & \textbf{Best practices} & \textbf{AL Citation\textsuperscript{$\ddagger$}} &\textbf{GL Citation\textsuperscript{$\ddagger$}} & \textbf{N} \\
	\hline

	Test Automaton Strategy & 
	    Define an effective test automation strategy & 
	    Eva: \citeP[P][]{Srodrigues2016relevance}, \citeP[P][]{Sgarousi2020exploring}
	    \newline 
	    Exp: \citeP[P][]{Skoomen2013tmap}, \citeP[P][]{furtado}, \citeP[P][]{Spersson2004establishment}, \citeP[P][]{Skoomen1999test}, \citeP[P][]{SsoftwareTestAutomation}, \citeP[P][]{Skhan2018issues} \newline 
	    Other: \citeP[P][]{Sgarousi2016}, \citeP[P][]{Sgarousi2017}, \citeP[P][]{Swiklund2017impediments}, \citeP[P][]{Swang2020survey},
	    \citeP[P][]{Sricca2021web}
	    
	    &  
	    Exp: \citepP[P][]{SGuru99}, \citeP[P][]{Sstbox3} , \citeP[P][]{S7BestPracticesinTestAutomation}, \citeP[P][]{SYang}, \citeP[P][]{Sskip}, \citeP[P][]{STATCOE-Report}, \citeP[P][]{SMichael}, \citeP[P][]{SThe15essentialbuildingblocks}, \citeP[P][]{SLessonsinTestAutomation}, \citeP[P][]{STestSPICE3}, \citeP[P][]{SjourneytoTestAutomationMaturity}, \citeP[P][]{SApire}, \citeP[P][]{SInfosys}, \citeP[P][]{SJoe2017}, \citeP[P][]{SUdi},
	    \citeP[P][]{SGryka2017},
	    \citeP[P][]{SAgie},
	    \citeP[P][]{SMatt2017},
	    \citeP[P][]{SAltexsoft}
	    \newline Opi: \citeP[P][]{SPavan}, 
	    \citeP[P][]{ScomplexCriticalSoftwareSysstemTesting},
	    \citeP[P][]{SCrombie},
	    \citeP[P][]{SAlan},
	    \citeP[P][]{SMatthews} & 37
         \\ \cline{2-5}
         & 
         Involve key stakeholders in strategy development & 
          Exp: \citeP[P][]{SsoftwareTestAutomation}, \citeP[P][]{Skoomen2013tmap}
         \newline Other: \citeP[P][]{Swiklund2017impediments} &  Exp: \citeP[P][]{Sstbox3}, \citeP[P][]{SInfosys}, \citeP[P][]{SCrombie} & 6
         \\ \cline{2-5}
         &
         Adjust the strategy to the changes& - &  
         Exp: \citeP[P][]{STestSPICE3}, \citeP[P][]{Sstbox3}, \citeP[P][]{SWQR17}, \citeP[P][]{SYan}, \citeP[P][]{SMichael} 
         \newline Opi: \citeP[P][]{SMatthews}& 6\\ \hline
         
    	Resources & 
    	Provide enough resources &
    	Eva: \citeP[P][]{Srodrigues2016relevance}
    	\newline
    	Exp: \citeP[P][]{SsoftwareTestAutomation},
    	\citeP[P][]{Skoomen2013tmap},
    	\citeP[P][]{Skoomen1999test},
    	\citeP[P][]{Spersson2004establishment}
    	\newline 
    	Other: \citeP[P][]{Skasurinen2010software}, 
    	\citeP[P][]{Sgarousi2017},
    	\citeP[P][]{Swiklund2017impediments},
    	\citeP[P][]{Swang2020survey}
    	& Exp:\citeP[P][]{SMichael},
    	\citeP[P][]{Sskip}, 
    	\citeP[P][]{SWQR17},
    	\citeP[P][]{STATCOE-Report},
    	\citeP[P][]{SYang},
    	\citeP[P][]{SMitchel},
    	\citeP[P][]{STestSPICE3}& 16 \\\hline

	Test organization & 
 	 Acquire enough management support for test automation &
 	  Exp: \citeP[P][]{Spersson2004establishment},
 	   \citeP[P][]{SsoftwareTestAutomation}
 	  \newline Other: \citeP[P][]{Sgarousi2016}, 
 	   \citeP[P][]{Slee2012survey},
 	   \citeP[P][]{Swiklund2017impediments},
 	   \citeP[P][]{Sdeak2013organization}& 
 	  Exp: \citeP[P][]{SthreeKeystoTestAutomation} & 7 \\ \cline{2-5}
 	 
	& Keep test professionals motivated about test automation &
	 Other: \citeP[P][]{Sdeak2013organization},
	\citeP[P][]{Swiklund2017impediments},
	\citeP[P][]{Slee2012survey} &
	Exp: \citeP[P][]{SthreeKeystoTestAutomation},
	\citeP[P][]{SLessonsinTestAutomation} & 5 \\\cline{2-5}
	
	& Have competent test professionals &
	Eva: \citeP[P][]{Srodrigues2016relevance} \newline
	Exp: \citeP[P][]{Skoomen2013tmap},
	\citeP[P][]{Swang2020test}\newline
	Other: \citeP[P][]{Sgarousi2017},
	\citeP[P][]{Sgarousi2016},
	\citeP[P][]{Swiklund2017impediments},
	\citeP[P][]{Swang2020survey} & 
	Exp: \citeP[P][]{Sstbox3},
	\citeP[P][]{SYang}, 
	\citeP[P][]{SLessonsinTestAutomation},
	\citeP[P][]{SWQR17},
	\citeP[P][]{STestSPICE3},
	\citeP[P][]{STATCOE-Report},
	\citeP[P][]{SDavid},
	\citeP[P][]{SjourneytoTestAutomationMaturity},
	\citeP[P][]{SMitchel},
	\citeP[P][]{SJoe2017},
	\citeP[P][]{SMaxim},
	\citeP[P][]{SAgie}; \newline
	Opi: \citeP[P][]{ScomplexCriticalSoftwareSysstemTesting},
	\citeP[P][]{SAlisha},
	\citeP[P][]{SCrombie} & 22\\ \cline{2-5}
	
	& Promote collaboration & Other: \citeP[P][]{Sdeak2013organization},
	\citeP[P][]{SWQR17}
	& 
	Exp: \citeP[P][]{Sstbox3},
	\citeP[P][]{SStackify},
	\citeP[P][]{SthreeKeystoTestAutomation},
	\citeP[P][]{StestDataManagmentMaturityAssessment},
	\citeP[P][]{STestSPICE3},
	\citeP[P][]{SPractiTest},
	\citeP[P][]{SUdi},
	\citeP[P][]{SLessonsinTestAutomation},
	\citeP[P][]{SAltexsoft}
	\newline 
	Opi: \citeP[P][]{SAlisha}
	& 12 \\\hline

	Knowledge \newline transfer &
    Share available test automation knowledge &
    Exp: \citeP[P][]{Skoomen1999test},
    \citeP[P][]{Skoomen2013tmap}
    \newline
    Other: \citeP[P][]{Sdeak2013organization}
    & 
    Exp: \citeP[P][]{SThe15essentialbuildingblocks},
    \citeP[P][]{SCentryLink},
    \citeP[P][]{SYang},
    \citeP[P][]{SLessonsinTestAutomation},
    \citeP[P][]{STATCOE-Report},
    \citeP[P][]{SCh4n},
    \citeP[P][]{SUladzislau} & 10 \\\cline{2-5}
    
	& Allow time for training and the learning curve &
	Other: \citeP[P][]{Slee2012survey},
	\citeP[P][]{Sgarousi2017},
	\citeP[P][]{Sgarousi2016},
	\citeP[P][]{Swiklund2017impediments},
	\citeP[P][]{Sdeak2013organization}
	& Exp: \citeP[P][]{SRanorex},
	\citeP[P][]{SJoe2017},
	\citeP[P][]{STATCOE-Report},
	\citeP[P][]{SUdi},
	\citeP[P][]{SLessonsinTestAutomation},
	\citeP[P][]{SMitchel},
	\citeP[P][]{Sstbox3}
	 \newline 
	Opi: \citeP[P][]{ScomplexCriticalSoftwareSysstemTesting}& 13\\ \hline

	Test tools & 
	 Select the right test tools & 
	 Eva: \citeP[P][]{Srodrigues2016relevance},
	 \citeP[P][]{Sborjesson2012},
	 \citeP[P][]{raulamo2019},
	 \citeP[P][]{Skasoju2013},
	 \citeP[P][]{Silles},
	 \citeP[P][]{Shaugset2011}
	 \newline
	 Exp: \citeP[P][]{SsoftwareTestAutomation},
	 \citeP[P][]{Spersson2004establishment},
	 \citeP[P][]{Skhan2018issues},
	 \citeP[P][]{Swang2020test},
	 \citeP[P][]{SGuru99}
	 \newline
	 Other: \citeP[P][]{Swiklund2017impediments},
	 \citeP[P][]{Sgarousi2016},
	 \citeP[P][]{Swang2020survey}
	 &
	  Exp: \citeP[P][]{ SLessonsinTestAutomation},
	 \citeP[P][]{SYang},
	 \citeP[P][]{ S7BestPracticesinTestAutomation},
	 \citeP[P][]{SYan},
	 \citeP[P][]{STOMsurvey},
	 \citeP[P][]{STATCOE-Report},
	 \citeP[P][]{SPlanningforLongtermTestAutomationSucess},
	 \citeP[P][]{SAtkins},
	 \citeP[P][]{SPractiTest},
	 \citeP[P][]{SWQR17},
	 \citeP[P][]{SWQR19},
	 \citeP[P][]{SWQR20}
	 \newline
	 Opi: \citeP[P][]{SAlisha},
	 \citeP[P][]{SBose}& 28\\ \cline{2-5}

	&  Properly use test tools &
	Exp: \citeP[P][]{Skoomen2013tmap},
	\citeP[P][]{SsoftwareTestAutomation}
	\newline 
	Other: \citeP[P][]{Slee2012survey},
	\citeP[P][]{Swiklund2012}
	& Exp: \citeP[P][]{SLessonsinTestAutomation},
	\citeP[P][]{SUdi},
	\citeP[P][]{SUladzislau} &7 \\\hline

	Test environment & 
  
     Set up good test environments &
     Eva: \citeP[P][]{Skasoju2013} \newline 
     Exp: \citeP[P][]{Skoomen2013tmap},
     \citeP[P][]{SsoftwareTestAutomation}
     \newline
     Other: \citeP[P][]{Swiklund2017impediments}, \citeP[P][]{Swang2020survey}
     &  Exp: \citeP[P][]{Sstbox3},
     \citeP[P][]{SThe15essentialbuildingblocks},
     \citeP[P][]{STATCOE-Report},
     \citeP[P][]{ SPlanningforLongtermTestAutomationSucess},
     \citeP[P][]{STestSPICE3},
     \citeP[P][]{SWQR17},
     \citeP[P][]{Sstbox3},
     \citeP[P][]{SRanorex},
     \citeP[P][]{STOMsurvey},
     \citeP[P][]{SJoe2017},
     \citeP[P][]{SLessonsinTestAutomation}
     \newline Opi: \citeP[P][]{SCrombie},
     \citeP[P][]{SMaxim},
     \citeP[P][]{SJason}& 19 \\ \cline{2-5}

    & Create high-quality test data & 
    Other: \citeP[P][]{Sricca2021web}
    & 
    Exp: \citeP[P][]{StestDataManagmentMaturityAssessment},
    \citeP[P][]{SWQR17},
    \citeP[P][]{S7BestPracticesinTestAutomation},
    \citeP[P][]{STestSPICE3},
    \citeP[P][]{STATCOE-Report},
    \citeP[P][]{SBrad}
    \newline
    Opinion: \citeP[P][]{SAlisha} & 8 \\ \hline
	
	Test automation requirements & 
	Define test automation requirements & 
	Exp: \citeP[P][]{Skoomen1999test}
	\newline
	Other: \citeP[P][]{Swang2020survey}
	& Exp: \citeP[P][]{STATCOE-Report},
	\citeP[P][]{SThe15essentialbuildingblocks},
	\citeP[P][]{STestSPICE3},
	\citeP[P][]{SAltexsoft},
	\citeP[P][]{SWQR20}
	\newline Other: \citeP[P][]{SMaxim},
	\citeP[P][]{SApire}
	& 9 \\ \cline{2-5}
	
	& Have control over changes of test automation requirements &
	Exp: \citeP[P][]{SsoftwareTestAutomation}
	\newline
	Other: \citeP[P][]{SSri2014},
	\citeP[P][]{Swang2020survey}
	& Exp: \citeP[P][]{SYang},
	\citeP[P][]{STATCOE-Report},
	\citeP[P][]{SAltexsoft} & 6
	\\ \hline
	
	Test design 
	& Develop high-quality test scripts & 
	Exp: \citeP[P][]{SsoftwareTestAutomation},
	\citeP[P][]{Spersson2004establishment},
	\citeP[P][]{Skoomen2013tmap}
	\newline
	Opi: \citeP[P][]{SBose}
	\newline
	Other: \citeP[P][]{SSri2014},
	\citeP[P][]{Swiklund2012},
	\citeP[P][]{Swiklund2017impediments},
	\citeP[P][]{Sricca2021web}
	& 
	Exp: \citeP[P][]{Sstbox3},
	\citeP[P][]{STATCOE-Report},
	\citeP[P][]{SBrain},
	\citeP[P][]{SYan},
	\citeP[P][]{STestSPICE3},
	\citeP[P][]{SDavid},
	\citeP[P][]{SLessonsinTestAutomation},
	\citeP[P][]{SBasilios},
	\citeP[P][]{SThe15essentialbuildingblocks},
	\citeP[P][]{SAmir},
	\citeP[P][]{SUladzislau},
	\citeP[P][]{SAhmed},
	\citeP[P][]{SMatt2017} & 21  \\\cline{2-5}
	
	& Arrange testware in good architecture & 
	 Exp: \citeP[P][]{SsoftwareTestAutomation},
	\citeP[P][]{Spersson2004establishment}
	\newline
	Other: \citeP[P][]{Swiklund2017impediments},
	\citeP[P][]{Swang2020survey},
	\citeP[P][]{Sricca2021web}
	& Exp: \citeP[P][]{Sstbox3},
	\citeP[P][]{SCentryLink},
	\citeP[P][]{S7BestPracticesinTestAutomation},
	\citeP[P][]{SMeasurementofQualityoftheTesting},
	\citeP[P][]{SAhmed},
	\citeP[P][]{SBrain},
	\citeP[P][]{SBasilios},
	\citeP[P][]{Swang2020test},
	\citeP[P][]{SRanorex} & 14
	\\ \hline
	
	Test execution &  
	Prioritize automated tests for execution &
	Exp: \citeP[P][]{SsoftwareTestAutomation}
	& Exp: \citeP[P][]{STestSPICE3},
	\citeP[P][]{SCentryLink},
	\citeP[P][]{SRanorex},
	\citeP[P][]{SMaxim},
	\citeP[P][]{SPractiTest},
	\citeP[P][]{SUdi},
	\citeP[P][]{STATCOE-Report},
	\citeP[P][]{SBrad} & 9 \\\cline{2-5}
	
    & Automate pre-processing and post-processing &
    Exp: \citeP[P][]{SsoftwareTestAutomation} 
    &  Exp: \citeP[P][]{STATCOE-Report} & 2  \\ \hline
    
    Verdicts & 
    Automate test oracles &
    -
    & Exp: \citepP[P][]{STATCOE-Report},
    \citeP[P][]{SWQR20} & 2 \\\cline{2-5}
    
    & Analyze test automation efficiently and effectively &
    Exp: \citepP[P][]{SsoftwareTestAutomation},
    \citeP[P][]{Skoomen2013tmap},
    \citeP[P][]{Spersson2004establishment}; 
    \newline
    Other: \citeP[P][]{Swiklund2017impediments}
    &
    Exp:\citepP[P][]{STATCOE-Report},
    \citeP[P][]{Sstbox3},
    \citeP[P][]{SRanorex},
    \citeP[P][]{STestSPICE3},
    \citeP[P][]{S7BestPracticesinTestAutomation},
    \citeP[P][]{Sskip},
    \citeP[P][]{STestinginaDevOpsMindset}
     & 11 \\\cline{2-5}
    
    & Report useful test automation results to key stakeholders &
    Exp: \citepP[P][]{SsoftwareTestAutomation}
    \newline Other: \citepP[P][]{garousi2017test},
    \citeP[P][]{Swang2020survey}
    & Exp: \citepP[P][]{Sstbox3},
    \citeP[P][]{STestSPICE3}
    \newline 
    Opi: \citepP[P][]{SJason}& 6
    \\ \hline
    
     Measurements& 
	Use the right test automation metrics
	&  Exp: \citeP[P][]{SsoftwareTestAutomation}; 
	\newline
	Other: \citeP[P][]{Swang2020survey} 
	& Exp: \citeP[P][]{SSealights},
	\citeP[P][]{STATCOE-Report},
	\citeP[P][]{SPlanningforLongtermTestAutomationSucess},
	\citeP[P][]{Sstbox3},
	\citeP[P][]{STestinginaDevOpsMindset},
	\citeP[P][]{STestSPICE3}
	\newline 
	Opi: \citeP[P][]{SMatthews}, \citeP[P][]{SMeasurementofQualityoftheTesting} & 10 \\\hline
     
     SUT & Design the SUT for automated testability &
     Eva: \citeP[P][]{Sborjesson2012} 
     \newline
     Exp: \citeP[P][]{Spersson2004establishment} \newline
     Other: \citeP[P][]{Swiklund2017impediments},
     \citeP[P][]{Srodrigues2016relevance},
     \citeP[P][]{Skasurinen2010software},
     \citeP[P][]{Swang2020survey}
     & 
     Exp: \citeP[P][]{SBrain},
     \citeP[P][]{S7BestPracticesinTestAutomation},
     \citeP[P][]{SPlanningforLongtermTestAutomationSucess},
     \citeP[P][]{SGryka2017},
     \citeP[P][]{SPractiTest}
     \newline 
     Opi: \citeP[P][]{SMaxim}& 12 \\\hline
    
      Technology & Adopt new technologies & - &  Exp: \citeP[P][]{SWQR17},
      \citeP[P][]{SWQR20},
      \citeP[P][]{SWQR19}
      \newline 
      Opi: \citeP[P][]{SScott} & 4 \\

  \hline
\multicolumn{4}{l}{  \footnotesize{\textsuperscript{$\ddagger$} The categories Eva, Exp, Opi, and Other have been introduced in Section 3.5.}}

\end{longtable} }

\subsection{RQ2 - The advice on how to conduct proposed best practices}	
\label{sec:resutls_RQ2}
Using cited sources (Table \ref{tab:factors}), we further explain 26 test automation best practices and present the advice given on how to conduct them. 

\subsubsection{Test automation strategy related practices.}
\label{sec:resultStrategyDefine}
 \paragraph{Define an effective test automation strategy.} 37 sources (Table \ref{tab:factors}) suggested defining an effective test automation strategy to set action plans to conduct test automation within an organization. Academic studies  \citepP[P][]{Srodrigues2016relevance}\citeP[P][]{Sgarousi2020exploring}, which respectively surveyed 33 and 72 industry professionals, found that the substantial improvement of test automation maturity can be achieved by effective strategic planning. As the guidance in the literature, many prior studies advised what main topics an effective test automation strategy should cover. Table~\ref{tab:TAStrategy} summarizes these main topics. The detailed analysis on how to define these main topics in strategy development was given in several test maturity models (proposed by authorized organizations). TMap \citeP[P][]{Skoomen2013tmap} advised the main steps to define test automation goals: Consider expected outcomes,  Transfer expected outcomes into goals, and Link the goals with organizational policies and business needs. It also advised the standard risk analysis process (from Identify risks, Analyze the impact of risks, to Define controlling measures) and resource allocation process (from Identify demand test automation resources, Assign resources and re-assign as necessary, to Track the usage of resources) for test automation projects. STBox 3.0  \citeP[P][]{Sstbox3} advised several approaches to identify test automation scope, e.g., based on prioritized features, stakeholder expectations, the ease of automation of test cases, and clients' feedback. It also illustrated the example approach to estimate test automation effort based on team members' performance. TestSPICE~3.0 \citeP[P][]{STestSPICE3} introduced an approach of analyzing the costs and benefits of multiple executions of automated tests.

\begin{table}[h]
\centering
     \caption{ \label{tab:TAStrategy} Test automation strategy}
    \begin{tabular}{|p{2.4cm} | p{6.8cm}| p{2.4cm} | p{4cm} |}
    \hline
    Main topic & Description & AL citation\textsuperscript{$\ddagger$} & GL citation\textsuperscript{$\ddagger$} \\ \hline
    Goals & Define the specific goals of test automation for a short or/and long period of time, e.g., increase the product quality and user satisfaction, support frequent releases, reduce testing time, enhance the testing scale, minimize human efforts, accelerate feedback loop. & 
    Exp: \citeP[P][]{Skoomen2013tmap}, \citeP[P][]{SsoftwareTestAutomation} &  
    Exp: \citeP[P][]{Sstbox3}, \citeP[P][]{SThe15essentialbuildingblocks}, 
    \citeP[P][]{STestSPICE3},
    \citeP[P][]{SInfosys},
    \citeP[P][]{SApire},
    \citeP[P][]{SAltexsoft}
    \newline Opi: \citeP[P][]{SPavan}
     \\ \hline
    
    Test scope & 
    Make decisions on which parts of the test scope shall be automated to what degree and to which test level  & 
     Exp:
    \citeP[P][]{Skoomen2013tmap},
    \citeP[P][]{SsoftwareTestAutomation}   & 
    Exp: \citeP[P][]{SYang},
    \citeP[P][]{Sgarousi2016},
    \citeP[P][]{STestSPICE3},
    \citeP[P][]{Swiklund2017impediments},
    \citeP[P][]{SMichael},
    \citeP[P][]{SApire},
    \citeP[P][]{STATCOE-Report},
    \citeP[P][]{Sstbox3},
    \citeP[P][]{SAltexsoft}\\ \hline
    
    Risks &
    Predict and track risks associated with test automation  & Exp: \citeP[P][]{Skoomen2013tmap},
    \citeP[P][]{Spersson2004establishment}
    \newline Other: \citeP[P][]{Swiklund2017impediments} &  
    Exp: \citeP[P][]{S7BestPracticesinTestAutomation},
    \citeP[P][]{Sstbox3},
    \citeP[P][]{SAltexsoft}
    \newline 
    Opi: \citeP[P][]{SPavan}\\ \hline
    
    Resources &  Consider what resources are required and available for test automation  & 
    Exp: \citeP[P][]{Skoomen2013tmap}, \citeP[P][]{SsoftwareTestAutomation} & 
    Exp: \citeP[P][]{STATCOE-Report},
    \citeP[P][]{SLessonsinTestAutomation},
    \citeP[P][]{SThe15essentialbuildingblocks},
    \citeP[P][]{Sstbox3},
    \citeP[P][]{SInfosys}
    \newline 
    Opi: \citeP[P][]{ScomplexCriticalSoftwareSysstemTesting} \\ \hline
    
    Costs \& benefits & 
    Analyze the benefits of test automation over its costs  &
    Exp: \citeP[P][]{Sgarousi2016},
    \citeP[P][]{Sgarousi2017}
    \newline Other:\citeP[P][]{SsoftwareTestAutomation}, \citeP[P][]{Spersson2004establishment} & 
    Exp: \citeP[P][]{STATCOE-Report},
    \citeP[P][]{SjourneytoTestAutomationMaturity},
    \citeP[P][]{SYang},
    \citeP[P][]{STATCOE-Report},
    \citeP[P][]{SInfosys}
    \newline 
    Opi: \citeP[P][]{SApire}, \citeP[P][]{SCrombie} \\ \hline
     
    Effort & Estimate how many efforts will be spent on test automation based on the defined strategy  & 
   Exp: \citeP[P][]{Skoomen2013tmap}, \citeP[P][]{SsoftwareTestAutomation} & 
   Exp: \citeP[P][]{STATCOE-Report} \newline 
    Opi: \citeP[P][]{ScomplexCriticalSoftwareSysstemTesting}, \citeP[P][]{SApire} \\ 
    \hline
    
    \multicolumn{4}{l}{  \footnotesize{\textsuperscript{$\ddagger$} The categories Eva, Exp, Opi, and Other have been introduced in Section 3.5.}}
    \end{tabular}
\end{table}

\paragraph{Involve key stakeholders in strategy development.} 6 sources (Table \ref{tab:factors}) noted the importance of involving key stakeholders (e.g., testers, developers, managers, and others who can affect or be affected by test automation) in strategy development and translate their expectations into meaningful decisions. Based on experience studies  \citeP[P][]{SCrombie}\citeP[P][]{SInfosys}\citeP[P][]{SsoftwareTestAutomation}, without key stakeholders' involvement, an organization may fail to identify important test automation needs, disappoint stakeholders, and miss out targets to make inputs. Referring to how to involve key stakeholders in strategy development, test maturity models TMap \citeP[P][]{Skoomen2013tmap} and STBox 3.0 \citeP[P][]{Sstbox3} suggested the formal approach - managers lead in working with key stakeholders to formally discuss the main topics on a test automation strategy - for the general software development contexts. Several practitioners \citeP[P][]{SCrombie}\citeP[P][]{SInfosys} viewed that, based on their experience, in agile contexts, it can be done in an informal approach, which emphasizes quick decisions and “agility in doing” (reactive planning) in responding to changing needs of key stakeholders. With such an informal approach, the main topics on a test automation strategy can be aware of at any time and informal discussion among key stakeholders will occur when necessary  \citeP[P][]{SCrombie}\citeP[P][]{SInfosys}. Yet, we did not found evaluation studies that validate or compare the effect of formal approach and informal approach in real practices.

\paragraph{Adjust the strategy to the changes.}  6 sources (Table \ref{tab:factors}) suggested adjusting the strategy to the changes, as they argued that change is the one true constant in test automation and it can occur in diverse forms, e.g., the technology, test scope, or business scope changes, test environment or SUT updates, the use of new test approaches, staff turnover. Practitioners \citeP[P][]{SYan}\citeP[P][]{SMichael} reported that, in their past projects, having a test automation strategy adjusted against the changes helped them to rethink their test automation needs and set up stepwise maturity improvements. As the guidance, test maturity model STBox 3.0 \citepP[P][]{Sstbox3} advised the main steps for adjusting the strategy against changes: Review the strategy around main topics it includes, Make changes with key stakeholders, and Communicate the changes within an organization. The detailed analysis for each step can be found in STBox 3.0 \citepP[P][]{Sstbox3}.

\subsubsection{Resources related practices.}  

\paragraph{Provide enough resources.}  16 sources (Table \ref{tab:factors}) proposed this best practice. Based on an industrial survey~\citeP[P][]{SWQR17} that explored the state of software testing practices with around 2000 test professionals in the world, the major obstacle for improving test automation maturity is resource shortage. An academic study \citeP[P][]{Srodrigues2016relevance} that surveyed 33 industry professionals found that provide enough resources is a key success factor for improving test automation maturity. Many experience studies  \citepP[P][]{Sskip}\citeP[P][]{SsoftwareTestAutomation}\citeP[P][]{Spersson2004establishment} reported that, in strategy development,  identifying demand resources and rationally allocating available resources for usage can avoid the risk of resource shortage to some extent. This links to `Define en effective test automation strategy' related practices presented in Section \ref{sec:resultStrategyDefine}. To support in identifying demand test automation resources, the literature introduced different types of test automation resources, see Table \ref{tab:TAResources}. Some practitioners \citeP[P][]{SMichael}\citeP[P][]{SYang} claimed that, based on their experience, using test management tools with resource management functions is helpful for rational resource allocation. Test maturity model TPI \citeP[P][]{Skoomen1999test} declared that management skills of managers also can affect the situation of resource allocation. This links to `Have competent test professional' related practices presented in Section \ref{sec:testOrganizationFactors}.

\begin{table}[htpb]
\centering
     \caption{ \label{tab:TAResources} Resources }
    \begin{tabular}{|p{1.8cm} | p{8cm}| p{3cm} | p{2.8cm} |}
    \hline
    Type & Resources & AL citation\textsuperscript{$\ddagger$} & GL citation\textsuperscript{$\ddagger$} \\ \hline
    Test \newline environment & Software, hardware, networks, cloud, operating systems, test data, supporting tools to set up a test environment. & 
    Exp: \citeP[P][]{Spersson2004establishment},
    \citeP[P][]{SsoftwareTestAutomation}
    \newline
    Other: \citeP[P][]{SWQR17},
    \citeP[P][]{SWQR19}
    &  
     Exp: \citeP[P][]{SYang},
    \citeP[P][]{SMitchel},
    \citeP[P][]{Sstbox3} \\ \hline
    
    Test tools & Test tools to support and perform test automation activities.  &
    Exp: \citeP[P][]{Skoomen1999test},
    \citeP[P][]{SsoftwareTestAutomation}
    \newline
    Other: \citeP[P][]{SWQR17},
    \citeP[P][]{SWQR19} & 
    Exp: \citeP[P][]{Sstbox3} 
    \\\hline

    Human resources & Skilled people, e.g., tool experts, test managers, testers, consultants, test automation experts. &
    Eva: \citeP[P][]{Srodrigues2016relevance} \newline
    Exp: \citeP[P][]{Spersson2004establishment},
    \citeP[P][]{SsoftwareTestAutomation}\newline
    Other: \citeP[P][]{SWQR17} 
    & 
    Exp: \citeP[P][]{Sstbox3},
    \citeP[P][]{SYang},
    \citeP[P][]{SMichael}
    \\ \hline
    
    Costs & Development and upkeep costs of test automation. Direct costs: resource purchasing, software licensing, training, cloud and network services, and people hiring. Indirect costs: efforts spent on test automation. 
 & 
 Other: \citeP[P][]{Swiklund2017impediments},
 \citeP[P][]{Skasurinen2010software},
 \citeP[P][]{SWQR17} & 
 Exp: \citeP[P][]{STATCOE-Report},
 \citeP[P][]{Sskip}\\\hline
 
     \multicolumn{4}{l}{ \footnotesize{\textsuperscript{$\ddagger$} The categories Eva, Exp, Opi, and Other have been introduced in Section 3.5.}}
    \end{tabular}
\end{table}

\subsubsection{Test organization related practices.}
\label{sec:testOrganizationFactors} 

\paragraph{Acquire enough management support for test automation.} 7 sources (Table \ref{tab:factors}) argued that, adopting test automation can bring work pattern, technology, and organizational culture changes, while introducing such changes need enough management support  within organizations. The academic study \citeP[P][]{Spersson2004establishment} presented a case study of establishing automated regression testing in two projects at ABB (a pioneering technology company) - automated regression testing almost failed in the middle of these two projects, since inadequate management support was obtained to involve core team members and get resource provision. However, we did not identify the advice on how to acquire management support for test automation from cited sources (Table \ref{tab:factors}) of this best practice.

\paragraph{Keep test professionals motivated about test automation.}  5 sources (Table \ref{tab:factors}) claimed that the extent of motivation determines how much effort test professionals would spend on test automation and the use rate of test tools, and therefore, they suggest keeping test professionals motivated about test automation. As described by Fewster and Graham \citeP[P][]{SsoftwareTestAutomation}, who had more than 20-year test automation experience, ``\textit{the best automation tool in the world will not help test efforts, if your team resists using it}". Based on academic studies  \citeP[P][]{Sdeak2013organization}\citeP[P][]{Slee2012survey} that surveyed the state of software testing practices in different organizations, having unmotivated test professionals may bring negative outcomes, e.g., more turnover, less productivity, low test efficiency, narrow test automation scope, and dissatisfaction to test automation outcomes. However, we did not identify the advice on how to motivate test  professionals about test automation from cited sources (Table \ref{tab:factors}) of this best practice.

\paragraph{Have competent test professionals.} 
\label{par:competency} As reported in 22 sources (Table \ref{tab:factors}), test professionals need the competencies (a collection of knowledge, skills, and abilities) to perform test automation tasks \citeP[P][]{Skoomen2013tmap}\citeP[P][]{ScomplexCriticalSoftwareSysstemTesting}. The academic study \citeP[P][]{Srodrigues2016relevance} on surveying 33 industrial practitioners found that having competent test professionals is key to improve test automation maturity. Based on a SLR on the impediments of test automation maturity \citepP[P][]{Swiklund2017impediments}, in terms of lacking competencies of test professionals, there is a high risk for test automation failures, low testing quality, and costs and efforts overdraft. The literature advised what competencies test professionals should have in test automation practices. Table~\ref{tab:TACompetencies} summarizes these competencies by role type: managerial role and technical role. Based on cited sources in Table~\ref{tab:TACompetencies}, in a practice context, a test professional can have both managerial role and technical role. To develop test professionals' competencies, test maturity models TMap \citepP[P][]{Skoomen2013tmap} and TestSPICE 3.0 \citeP[P][]{STestSPICE3} suggested organizing training programs, sharing the expertise, and learning from experience and practice. These practices are related to knowledge transfer related practices in Section \ref{sec:knowledgeTransfer}. Experience studies \citeP[P][]{SLessonsinTestAutomation}\citeP[P][]{Spersson2004establishment}\citeP[P][]{SMatt2017} have confirmed the increased competencies of test professionals after organizing training programs. Besides, TMap \citepP[P][]{Skoomen2013tmap} also advised organizations to hire new experienced individuals, e.g., test automation experts, tool specialists, or consultants with required competencies. The positive effect of hiring new experienced individuals on accelerating competence development has been noticed by practitioners \citeP[P][]{SLessonsinTestAutomation}\citeP[P][]{Swang2020test} in past test automation practices.

\begin{table}[ht]
\centering
     \caption{Competencies of test professionals  \label{tab:TACompetencies} }
    \begin{tabular} {|p{1.7cm}|p{8.8cm}|p{2.4cm}|p{2.9cm}|}
    \hline
     & Test automation related competencies & AL citation\textsuperscript{$\ddagger$} & GL citation\textsuperscript{$\ddagger$} \\ \hline
    
    Managerial role & 
   Ability to develop, adjust, and execute the test automation strategy within planning and budget &
   Exp: \citeP[P][]{Skoomen2013tmap},
   \citeP[P][]{SWQR17} \newline
   Other: \citeP[P][]{Swiklund2017impediments} & 
   -\\ \cline{2-4}
   
   & Ability to coordinate test automation activities and stakeholders &
   
    Exp: \citeP[P][]{Skoomen2013tmap} \newline 
    Other: \citeP[P][]{Swiklund2017impediments} &
    - \\ \cline{2-4}
   
   & Knowledge and skills for test automation maturity improvement &
   - & 
   Exp: \citeP[P][]{SWQR17},
   \citeP[P][]{Sstbox3},
   \citeP[P][]{STATCOE-Report},
   \citeP[P][]{SMitchel}\\ \cline{2-4}
   
   & Knowledge of architectures and tools for system development
     & 
     Exp: \citeP[P][]{Skoomen2013tmap} \newline
     Other: \citeP[P][]{Swiklund2017impediments} 
     & 
     Exp: \citeP[P][]{SLessonsinTestAutomation},
     \citeP[P][]{SWQR17}\\ \hline
    
    Technical role & 
    Coding and scripting abilities & 
    Exp: \citeP[P][]{Skoomen2013tmap} \newline
    Other: \citeP[P][]{Swiklund2017impediments} 
    & 
    Exp: \citeP[P][]{STATCOE-Report},
    \citeP[P][]{SYang},
    \citeP[P][]{SWQR17},
    \citeP[P][]{SMitchel},
    \citeP[P][]{SWQR17}
    \newline 
    Opi: \citeP[P][]{ScomplexCriticalSoftwareSysstemTesting}\\ \cline{2-4}
    
    & Expertise with test tools and the capability to use them &
     Other: \citeP[P][]{Swiklund2017impediments}
     & 
      Exp: \citeP[P][]{SWQR17},
     \citeP[P][]{Sstbox3},
     \citeP[P][]{SLessonsinTestAutomation}  \\ \cline{2-4}
     
    & Knowledge about the SUT &
    Other: \citeP[P][]{Swiklund2017impediments}  & 
    Exp: \citeP[P][]{SMitchel}, 
    \citeP[P][]{SWQR17} \newline
    Opi: \citeP[P][]{ScomplexCriticalSoftwareSysstemTesting}\\ \cline{2-4}
    & Skills in creating and interpreting the test basis (e.g., requirements, test results, functional and technical design) & - &
    Exp: \citeP[P][]{SjourneytoTestAutomationMaturity},
    \citeP[P][]{SWQR17},
    \citeP[P][]{SMitchel} \\ \cline{2-4}
    
    & Ability to automate tests, support and maintain test artifacts, prepare and maintain test environment and test data, and analyze test automation results.     
     & 
     Exp: \citeP[P][]{Skoomen2013tmap} \newline
     Other: \citeP[P][]{Swiklund2017impediments} &
     Exp: \citeP[P][]{SjourneytoTestAutomationMaturity},
     \citeP[P][]{SMitchel},
     \citeP[P][]{SWQR17}\\ \hline

    \multicolumn{4}{l}{  \footnotesize{\textsuperscript{$\ddagger$} The categories Eva, Exp, Opi, and Other have been introduced in Section 3.5.}}
    \end{tabular}
\end{table}

\paragraph{Promote collaboration.} 12 sources (Table \ref{tab:factors}) declaimed that, test automation is a team effort that needs many individuals working together to reach the common goals, so that promoting collaboration in test automation practices is suggested. Based on the World Quality report 2017-18 \citeP[P][]{SWQR17}, good collaboration can ensure that test automation efforts are distributed evenly, the competency of everyone is fully used, and the roles and responsibilities of individuals are clearly defined. The experience of practitioners  \citeP[P][]{SStackify}\citeP[P][]{SthreeKeystoTestAutomation} showed that, by collaborating more, people can better share knowledge and learn from others. However, from cited 12 sources (Table \ref{tab:factors}), we did not identify the advice on how to promote collaboration in test automation practices.

 \subsubsection{Knowledge transfer related practices}
 \label{sec:knowledgeTransfer}
 \paragraph{Share available test automation knowledge.}
10 sources (Table \ref{tab:factors}) suggested sharing available test automation knowledge to reduce duplicated efforts, encourage sharing and learning from others, and collect knowledge for further reference and innovations. Based on academic studies  \citeP[P][]{Sdeak2013organization}\citeP[P][]{SSri2014} that surveyed different organizations about the state of their software testing practices, the result of not sharing test automation knowledge is the drain of knowledge, decreased innovation, reduced efficiency, and increased test automation costs. To provide guidance on sharing test automation knowledge, prior studies introduced different types of shareable test automation knowledge and also advised the general places to share it, as shown in Table \ref{tab:TAKnowledge}. The details on how to share these types of test automation knowledge through advised places were not identified from cited sources in Table \ref{tab:TAKnowledge}.

 \begin{table}[h]
\centering
     \caption{ \label{tab:TAKnowledge} Knowledge sharing }
    \begin{tabular}{|p{2cm} | p{8cm}| p{2.7cm} | p{2.7cm}|}
    \hline
    & Resources & AL citation\textsuperscript{$\ddagger$} & GL citation\textsuperscript{$\ddagger$} \\ \hline
    Knowledge type & Test scripts, test cases, testing techniques, testing approaches, test tool expertise, test automation requirements, test automation metrics, test automation expertise, development guidelines, system requirements, test reports, best practices and lesson learned.  &  
    Exp: \citeP[P][]{Skoomen2013tmap},
    \citeP[P][]{Skoomen1999test}
    & 
    Exp: \citeP[P][]{SYang},
    \citeP[P][]{STATCOE-Report},
    \citeP[P][]{SLessonsinTestAutomation},
    \citeP[P][]{SThe15essentialbuildingblocks}\\ \hline

     Place to share & Wikis, annotations, content management systems, internal/external open source repository, group discussion forums, instant messaging tools, networking platforms, training sessions. & 
     Exp: \citeP[P][]{Skoomen2013tmap},
     \citeP[P][]{SCentryLink},
     \citeP[P][]{SYang},
     \citeP[P][]{SLessonsinTestAutomation}
     \newline
    Other: \citeP[P][]{Sdeak2013organization} & 
    Exp: \citeP[P][]{SUladzislau}\\ \hline
    
    \multicolumn{4}{l}{  \footnotesize{\textsuperscript{$\ddagger$} The categories Eva, Exp, Opi, and Other have been introduced in Section 3.5.}}
    \end{tabular}
\end{table}

\paragraph{Allow time for training and learning curve.} \label{par:training} 
13 sources (Table \ref{tab:factors}) suggested organizing training to develop the competencies of test professionals in test automation practices. What competencies test professionals should have are presented in Table \ref{tab:TACompetencies} (Section \ref{sec:testOrganizationFactors}). As reported in Ranorex's technical report~\citeP[P][]{SRanorex}, based on the experience of many test automation consultants, training provides both organizations and individuals as a whole with benefits: organizations offer consistent test automation knowledge and development guidelines to individuals through training; By taking training, individuals are able to develop the required competencies and thus increase the performance in the current role. Many experience studies \citeP[P][]{SJoe2017}\citeP[P][]{SUdi}\citeP[P][]{SLessonsinTestAutomation}\citeP[P][]{SMitchel} reported that, the learning curve to develop competencies is different from individual to individual in practice- if the learning time needed for test automation is underestimated or reduce, individuals may suffer the difficulty to complete the steep learning curve and then become less productive and motivated for test automation. However, from cited sources in Table \ref{tab:factors}, we did not identify the advice on how to organize training and estimate/improve individuals' learning curves in test automation practices.

\subsubsection{Test tools related practices.} 
\label{sec:testtools}
\paragraph{Select the right test tools.} 
28  sources (Table \ref{tab:factors}) recommended selecting the right test tools.  According to the recent World Quality Reports \citeP[P][]{SWQR19} \citeP[P][]{SWQR20} (published by Micro Focus, Capgemini, and Sogeti), test tools available in the market vary in different functions, price, learning curve, configuration settings, attributes (e.g., usability, extendability, maintainability, etc), programming language, future development trends.  Several scholars \citeP[P][]{Srodrigues2016relevance}\citeP[P][]{raulamo2019}\citeP[P][]{Silles} have surveyed test professionals from different organizations and evaluated that, the right test tools refer to ``test tools that can best fit the current test automation needs of an organization''. To guide organizations to select the right test tools, the literature has proposed test tool selection criteria and we conclude that in Table \ref{tab:TATesttoolselecton}. The content validity of each criterion in Table \ref{tab:TATesttoolselecton} has been evaluated by a recent academic study \citeP[P][]{raulamo2019} that surveyed 89 test professionals in the industry. Compared to \citeP[P][]{raulamo2019}, academic studies \citeP[P][]{Sborjesson2012}\citeP[P][]{Silles} that surveyed fewer number of test professionals also evaluated the content validity of some criteria. Yet, the conflict view was presented in an experience study \citeP[P][]{Swang2020test} that reports the 10-year experience of test automation maturity improvement in a small size DevOps team. It described that, selecting the right test tools based on pre-defined criteria is less useful than selecting each test tool with an experimentation mindset, because new needs that offend pre-defined criteria may emerge at anytime.

{
\begin{longtable}{|p{2.3cm} | p{8.4cm}| p{2.65cm} | p{2.65cm}|} 
     \caption{ \label{tab:TATesttoolselecton} Test tool selection criteria}
    \\\hline 
    Criterion & Description &  AL citation\textsuperscript{$\ddagger$} &  GL citation\textsuperscript{$\ddagger$} \\ \hline

    Function & Consider which testing functions are currently required, e.g., unit testing, performance testing, functional testing, acceptance testing, GUI testing, non-functional testing. & 
    Eva: \citeP[P][]{Silles},
    \citeP[P][]{raulamo2019} \newline 
    Exp: \citeP[P][]{SsoftwareTestAutomation}  & 
    Exp: \citeP[P][]{SYang}, \citeP[P][]{SGuru99} \\ \hline
    
    Feature & Identify which features of a test tool are currently required, e.g., can be integrated with continuous integration server or/and bug tracking systems, uses scriptless test automation, provides test design patterns, supports test execution and logging, has a real-time dashboard and simple reporting.  & 
      Eva: \citeP[P][]{raulamo2019}, \citeP[P][]{Silles} &  Exp: \citeP[P][]{S7BestPracticesinTestAutomation}, \citeP[P][]{SYan},
     \citeP[P][]{SGuru99}\\ \hline
    
    Usability & Assess whether a test tool is easy to use, its training time is short, and its learning curve is low.   &  
    Eva: \citeP[P][]{raulamo2019},
    \citeP[P][]{Silles} & 
     Exp: \citeP[P][]{SGuru99},
     \citeP[P][]{Sborjesson2012},
     \citeP[P][]{SsoftwareTestAutomation}\\ \hline
    
    Maintainability  & Investigate how easy a test tool can be maintained in the current organization & 
    Eva: \citeP[P][]{raulamo2019}, 
    \citeP[P][]{Silles}
    \newline 
    Exp: \citeP[P][]{SsoftwareTestAutomation}& 
    Exp: \citeP[P][]{SYang}\\ \hline
    
    Compatibility & Consider the compatibility with the SUT and the current system architecture. & 
    Eva: \citeP[P][]{raulamo2019} &  
    Exp: \citeP[P][]{SYang},
    \citeP[P][]{SLessonsinTestAutomation},
    \citeP[P][]{S7BestPracticesinTestAutomation}\\\hline 
    
    Costs \& benefits & Analyze whether the potential benefits of buying a test tool would outweigh the costs. &  Eva: \citeP[P][]{raulamo2019} &  Exp: \citeP[P][]{SLessonsinTestAutomation},
    \citeP[P][]{SAlisha},
    \citeP[P][]{Silles}\\\hline

    Environmental constraints &  Explore the environmental constraints that affect the use of a test tool, e.g., a test tool is specific to particular hardware, software, SUT, operating systems, and programming languages supported. & 
    Eva: \citeP[P][]{Sborjesson2012},
    \citeP[P][]{Silles},
    \citeP[P][]{raulamo2019}
    \newline
    Exp:  \citeP[P][]{SsoftwareTestAutomation} &  
    Exp: \citeP[P][]{SYang},
    \citeP[P][]{STATCOE-Report}
    \newline 
    Opi: \citeP[P][]{SAlisha}\\ \hline
    
    Organizational constrains & Understand organizational constraints that affect the use of a test tool, e.g., available resources, the budget, organizational policy, the innovation roadmap, and the development strategy of an organization. & 
    Eva: \citeP[P][]{Silles},
    \citeP[P][]{raulamo2019}
    \newline
    Exp: \citeP[P][]{SsoftwareTestAutomation} & 
    - \\ \hline
    
    Licensing & Examine the licensing type of a test tool, e.g., open source or commercial licensing. &
    Eva: \citeP[P][]{Sborjesson2012},
    \citeP[P][]{Silles},
    \citeP[P][]{raulamo2019} &
    Exp: \citeP[P][]{SYang}, \citeP[P][]{Sborjesson2012} \\ \hline
    
    Vendor support & Check what support the tool vendor provides, e.g., availability of training, user community, warranty, installation, upgrades, help services.  & 
    Eva: \citeP[P][]{Sborjesson2012},
    \citeP[P][]{Silles},
    \citeP[P][]{raulamo2019} & 
    Exp: \citeP[P][]{STATCOE-Report},
    \citeP[P][]{SPlanningforLongtermTestAutomationSucess},
    \citeP[P][]{Silles} \\ \hline
    
    \multicolumn{4}{l}{  \footnotesize{\textsuperscript{$\ddagger$} The categories Eva, Exp, Opi, and Other have been introduced in Section 3.5.}}
\end{longtable}}

 \paragraph{Properly use test tools.} 
7 sources (Table \ref{tab:factors}) indicated that the carefully selected test tools may not deliver the promised outcomes, if they are not properly used. The specific example was illustrated in Lee~et al.'s academic study~\citeP[P][]{Slee2012survey} that surveyed the industry about the usage of test tools in practice. Their study results showed that, in sampled 73 organizations, defect management tools were not functioning as expected, since the use of these tools was limited to basic functions (such as defect reporting, recording, and tracking) but seldom for more complex functions (like defect prediction and prevention). As reported by many practitioners \citeP[P][]{SLessonsinTestAutomation}\citeP[P][]{SUdi}\citeP[P][]{SUladzislau}, based on their experience, properly using test tools requires sufficient test automation knowledge and tool expertise.  This can be linked to ``Have competent test professionals" and ``Allow for training and learning curve" related practices presented in Section \ref{par:competency} and Section \ref{sec:knowledgeTransfer} respectively.

\subsubsection{Test environment related practices.}

\paragraph{Set up good test environments.} 
 19 sources (Table \ref{tab:factors}) suggested setting up good test environments to improve test automation maturity. As defined by the software industry and research community, test environment refers to a context that consists of software, hardware, devices, network, storage, test data, servers, and other facilities to execute automated tests \citeP[P][]{SWQR17}\citeP[P][]{SWQR19}\citeP[P][]{SWQR20}. Based on many experience studies  \citeP[P][]{SsoftwareTestAutomation}\citeP[P][]{SJoe2017}\citeP[P][]{Sstbox3}\citeP[P][]{SRanorex}, poor test environments may cause the lack of supporting facilities, unstable tests, test execution failures, and thus the loss of testing time and costs in test automation practices. To help the organizations to assess  test environments, the literature attempted to characterize the good test environment. Table \ref{tab:TATestEnvrionment} collects the good test environment's characteristics from different sources. Besides, several studies advised approaches to set up good test environments and improve existing test environments, as summarized in Table \ref{tab:TATestEnvrionmentAppraoch}. Technical reports \citeP[P][]{STATCOE-Report}\citeP[P][]{SWQR17} (from software organizations) mentioned that test environment automation has many benefits: increase the agility to configure/re-configure test environments, ease to defect the conflicts of configured artifacts, reduce the operation costs, and avoid human errors. They recommended test environment automation for test automation on a large scale, in a continuous integration context, and has multiple versions of test environments. Other approaches (presented in Table \ref{tab:TATestEnvrionment}) were recommended by cited sources for the general test automation context.

\begin{table}[htpb]
\centering
     \caption{ \label{tab:TATestEnvrionment} Characteristics of the good test environment}
    \begin{tabular}{|p{2.4cm} | p{8cm}| p{2.65cm} | p{2.66cm} |}
    \hline
       & Characteristics &  AL citation\textsuperscript{$\ddagger$} &  GL citation\textsuperscript{$\ddagger$}\\ \hline
      Effectiveness & Close to the corresponding product environment &  - & 
       Exp: \citeP[P][]{SThe15essentialbuildingblocks}, 
       \citeP[P][]{SJoe2017},
       \citeP[P][]{SYang},
       \citeP[P][]{SPavan}\\ \hline
      
      Availability & Available all the times to run automated tests &  
      Other: \citeP[P][]{Swiklund2017impediments} & 
       Exp: \citeP[P][]{SPavan},
       \citeP[P][]{SDorothy2009},
       \citeP[P][]{SPlanningforLongtermTestAutomationSucess},
       \citeP[P][]{SWQR17}\\ \hline
      
      Stability & There is little chance for outage & 
      Other: \citeP[P][]{Swiklund2017impediments},
      \citeP[P][]{Sgarousi2016} &  
       Opi: \citeP[P][]{ScomplexCriticalSoftwareSysstemTesting} \\\hline
      
      Agility & Easy to be configured/re-configured  & 
      Other: \citeP[P][]{Swiklund2017impediments} & 
      Exp: \citeP[P][]{SJoe2017}, 
      \citeP[P][]{STATCOE-Report}\\ \hline
      
      Consistency & Run in an identical state for the same tests in different test phases &-& 
      Exp: \citeP[P][]{STATCOE-Report}, \citeP[P][]{SLessonsinTestAutomation} \\ \hline
      
      Reproducibility & Can quickly recover from the points of failures &
      Exp: \citeP[P][]{SsoftwareTestAutomation} &
      Exp: \citeP[P][]{SYang}, \citeP[P][]{SJoe2017} \\ \hline
      
      Maintainability & Easy to maintain and update & - & 
      Exp: \citeP[P][]{SWQR17}, \citeP[P][]{Sstbox3}
      \\\hline
      
    \multicolumn{4}{l}{  \footnotesize{\textsuperscript{$\ddagger$} The categories Eva, Exp, Opi, and Other have been introduced in Section 3.5.}}
    \end{tabular}
\end{table}
\vspace{-10mm}
\begin{table}[h]
\centering
     \caption{ \label{tab:TATestEnvrionmentAppraoch} Approaches to set-up/improve test environments}
    \begin{tabular}{ |p{3cm}| p{7cm} | p{2.65cm} | p{2.65cm} |}
    \hline
      Approaches & Description & AL citation\textsuperscript{$\ddagger$} & GL citation\textsuperscript{$\ddagger$} \\ \hline
      
      Production environment understanding & Comprehend the corresponding production environment & - & 
      Exp: \citeP[P][]{Sstbox3}, \citeP[P][]{SYang} \newline
      Opi: \citeP[P][]{SPavan}\\ \hline
      
      Requirement analysis & Analyze test automation requirements for configurations of test environments  & - & Exp: \citeP[P][]{SPlanningforLongtermTestAutomationSucess}, \citeP[P][]{SLessonsinTestAutomation}, 
      \citeP[P][]{Sstbox3} \\ \hline
      
      User analysis & Understand the demand and usage of test teams & - & Exp:\citeP[P][]{STATCOE-Report} \\ \hline 
      
      Environment testing & Test new test environments before actual usage & 
      Exp:\citeP[P][]{SsoftwareTestAutomation} \newline
      Other: \citeP[P][]{Swiklund2017impediments} & 
      -\\ \hline 
      
      Test environment \newline automation & Automate test environment configuration activities ( e.g., environment design, provision, deployment, operation) and the relevant supporting activities (e.g., planning, maintenance, self-healing) & - & 
      Exp: \citeP[P][]{STATCOE-Report}, \citeP[P][]{SWQR17} \\ \hline 
      
     \multicolumn{4}{l}{  \footnotesize{\textsuperscript{$\ddagger$} The categories Eva, Exp, Opi, and Other have been introduced in Section 3.5.}}
    \end{tabular}
\end{table}

 \paragraph{Create high-quality test data.} 
Referring to 8 cited sources (Table \ref{tab:factors}), test data quality affects test automation maturity. Based on the experience of practitioners  \citeP[P][]{StestDataManagmentMaturityAssessment}\citeP[P][]{SWQR17}\citeP[P][]{S7BestPracticesinTestAutomation}, low-quality test data can delivery wrong results, increase rework efforts, and waste testing time in test automation. To support organizations to assess the quality of test data, the literature defined the main characteristics of high-quality test data, see Table~\ref{tab:HighquailityTData}. Furthermore, prior experience studies (in the GL) presented several approaches to ensure/improve test data quality, while researchers' advice in the AL was not identified. Many experience studies advised setting up the guidelines for test data generation and management within an organization. Table \ref{tab:HighquailityTDataapproach} collects what guidelines these experience studies advised to set up for. Experience studies  \citeP[P][]{StestDataManagmentMaturityAssessment}\citeP[P][]{SWQR17} reported that, using test automation frameworks, which  have built-in standard guidelines for test data generation and management and are comprised of supporting practices and tools, can help to create high-quality test data. The experience study \citeP[P][]{StestDataManagmentMaturityAssessment} also suggested that manually reviewing the delivered test data for quality is also helpful.

 \begin{table}[h]
\centering
     \caption{ \label{tab:HighquailityTData}  High quality test data}
    \begin{tabular}{|p{2cm} | p{8cm}| p{2cm} | p{4cm}|}
    \hline
      Characteristic & Description & AL citation\textsuperscript{$\ddagger$} & GL citation\textsuperscript{$\ddagger$} \\ \hline
      
     Completeness & Contain all required information & - & 
     Exp: \citepP[P][]{Sstbox3}, \citeP[P][]{STestSPICE3} \\ \hline 
     
     Accuracy & Store the accurate information that reflects a real situation & - & 
     Exp: \citepP[P][]{Sstbox3}, \citeP[P][]{STestSPICE3}, 
     \citeP[P][]{SWQR17} \\ \hline
     
     Consistency & The piece of information does not contradict another piece of information & 
     Exp: \citepP[P][]{SsoftwareTestAutomation}& Exp:\citepP[P][]{STATCOE-Report} \\ \hline
     
     Usage & Easy to access and use & 
     Exp: \citeP[P][]{SsoftwareTestAutomation} &
     Exp: \citeP[P][]{STestSPICE3}, \citeP[P][]{SWQR17} \\ \hline
     
     Maintenance & Easy to maintain and make changes  &
     Exp: \citeP[P][]{SsoftwareTestAutomation} & 
     Exp: \citepP[P][]{STestSPICE3}, \citeP[P][]{SBrain}, \citeP[P][]{StestDataManagmentMaturityAssessment}, \citeP[P][]{SWQR17}  \\ \hline
    
     Reuse & Easy to reuse  &
     Exp: \citeP[P][]{SsoftwareTestAutomation} &
     Exp: \citepP[P][]{STestSPICE3}, \citeP[P][]{STATCOE-Report}
      \\ \hline
      
     Privacy & Compliance with data policies & - & 
     Exp: \citeP[P][]{StestDataManagmentMaturityAssessment}, \citeP[P][]{Sstbox3}, \citeP[P][]{SWQR17} \\ \hline
     
     \multicolumn{4}{l}{  \footnotesize{\textsuperscript{$\ddagger$} The categories Eva, Exp, Opi, and Other have been introduced in Section 3.5.}}
    \end{tabular}
\end{table}

 \begin{table}[htpb]
\centering
     \caption{ \label{tab:HighquailityTDataapproach}  Test data generation and management guidelines}
    \begin{tabular}{| p{6.4cm}| p{3.8cm}|}
    \hline
     Guideline & GL citation\textsuperscript{$\ddagger$} \\ \hline
     What test data are required & 
     Exp: \citeP[P][]{Sstbox3}, \citeP[P][]{STestSPICE3} \\\hline
     How to generate test data  & 
     Exp: \citeP[P][]{StestDataManagmentMaturityAssessment}, 
     \citeP[P][]{S7BestPracticesinTestAutomation} \\\hline
     
     How to use, manage, and maintain test data  & 
     Exp: \citepP[P][]{STestSPICE3}, \citeP[P][]{SBrad}\\\hline
     
     When and how to update test data & 
     Exp: \citeP[P][]{StestDataManagmentMaturityAssessment}, \citeP[P][]{STestSPICE3}\\\hline
     
     How to deal with legacy test data  & 
     Exp: \citepP[P][]{StestDataManagmentMaturityAssessment} \\\hline
     
     What data policies should follow  &
     Exp: \citeP[P][]{StestDataManagmentMaturityAssessment}, \citeP[P][]{Sstbox3}, 
     \citeP[P][]{SWQR17} \\\hline
    
    \multicolumn{2}{l}{  \footnotesize{\textsuperscript{$\ddagger$} The categories Eva, Exp, Opi, and Other have been introduced in Section 3.5.}}
    \end{tabular}
\end{table}

\subsubsection{Test automation requirements related practices.}
\label{sec:RQ2_results_requirements}
\paragraph{Define test automation requirements.} 
9 sources (Table \ref{tab:factors}) suggested defining test automation requirements to state expected test automation features and functionalities. Based on technical reports~\citeP[P][]{SAltexsoft}\citeP[P][]{STATCOE-Report} from software organizations, if test automation requirements are not defined, test automation may lose the right coverage and depth, consume development resources, and have low performance. Test maturity models TMap \citeP[P][]{Skoomen2013tmap} and TestSPICE 3.0 \citeP[P][]{STestSPICE3} suggested that, both test automation requirements and manual testing requirements are software testing requirements. They advised the example steps to define software testing requirements according to system requirements, test levels, stakeholder expectations, and product risks. The recent World Quality Reports \citeP[P][]{SWQR20}\citeP[P][]{SWQR19}  mentioned that test automation requirements are different from manual testing requirements. They noted that, in the current industry, many organizations are using machine learning techniques to automatically extract functional requirements from existing test cases (which supply inputs to test execution and outputs in response to inputs), but details on how they do that were not introduced.

 \paragraph{Have control over changes of test automation requirements}  
6 sources (Table \ref{tab:factors}) mentioned that test automation requirements may change throughout a testing process, e.g., when the major changes are made on the SUT, stakeholders alter expectations, business requirements change, and test automation goals shift, and hence, they recommended organizations to have control over changes of test automation requirements. Based on the experience of practitioners  \citeP[P][]{SYang}\citeP[P][]{SsoftwareTestAutomation}, if test automation requirements change at the early stage, communicating the changes with stakeholders and giving enough time to reset test automation may be easy; However, if the changes occur at the late stage, it would need lots of rework efforts. The advice on how to control changes of test automation requirements was given in the study \citeP[P][]{SsoftwareTestAutomation}, which  proposed test automation guidelines according to the authors' decades of test automation experience and field studies of case organizations. Based on this study~\citeP[P][]{SsoftwareTestAutomation}, the actions that can be taken to control the requirement changes include: Analyzing risks of requirement changes at the initial phase, Preparing flexible test cases to automate, Automating the tests which remain unchanged first, and Producing flexible and changeable test scripts in nature. The detailed analysis on each action is described in the same study \citeP[P][]{SsoftwareTestAutomation}. We found that,  the action ``Producing flexible and changeable test scripts'' links to ``Develop high-quality test scripts'' related practices in Section \ref{sec:ResultsTestDesign}.

\subsubsection{Test design related practices.}
\label{sec:ResultsTestDesign}

\paragraph{Develop high-quality test scripts.} 
Developing high-quality test scripts (written in test code) is important to mature test automation, as presented in 21 sources (see Table \ref{tab:factors}). Based on a SLR \citeP[P][]{Swiklund2017impediments} on the impediments of test automation maturity, low-quality test scripts may make automated tests to be prone to errors (e.g., security vulnerabilities, unreliable results, and coding standard violations) and difficult to run, maintain, and reuse. To help the organizations to assess test script quality, prior studies characterized the high-quality test scripts. Table \ref{tab:TestScriptC} collects the characteristics of high-quality test scripts from different sources. Several experience studies reported approaches to create or improve test scripts; Table \ref{tab:TestCodeQuality} presents these approaches. As reported in cited sources  \citepP[P][]{SsoftwareTestAutomation}\citeP[P][]{STestSPICE3}\citeP[P][]{Sstbox3} (Table \ref{tab:TestCodeQuality}), on the basis of some approaches, relevant supporting tools have been developed: scripting technique based tools, statistic code analysis tools, and test automation frameworks (provide standard development guidelines). Formal check and Peer review need human judgment  \citeP[P][]{SLessonsinTestAutomation}\citeP[P][]{STestSPICE3}.

\begin{table}[h]
\centering
     \caption{ \label{tab:TestScriptC} Characteristics of high quality test scripts}
    \begin{tabular}{|p{2cm} | p{7.8cm}| p{2.5cm} | p{3.6cm}|}
    \hline
      & Characteristic & AL citation\textsuperscript{$\ddagger$} & GL citation\textsuperscript{$\ddagger$} \\ \hline

     Size & Written in small size & 
     Exp: \citepP[P][]{SsoftwareTestAutomation} & 
     Exp: \citeP[P][]{STATCOE-Report} \\ \hline
     
     Function & Have the clear goal(s) to preform a single task & - & Exp: \citepP[P][]{SThe15essentialbuildingblocks}, \citeP[P][]{STestSPICE3} \\ \hline
     
     Structure & Have the clear and simple structure that are easy to understand  & 
     Exp:  \citeP[P][]{SsoftwareTestAutomation}& 
     Exp: \citeP[P][]{SBrain}, \citeP[P][]{STestSPICE3}, 
     \citeP[P][]{SMaxim} \\ \hline
     
     Maintenance & Easy to maintain and make changes  & 
     Exp: \citeP[P][]{SsoftwareTestAutomation} & 
    Exp: \citepP[P][]{STestSPICE3}, \citeP[P][]{SBrain}, 
    \citeP[P][]{SMaxim} \\ \hline
    
     Reuse & Easy to be reused by several test cases  & Exp: \citeP[P][]{SsoftwareTestAutomation}& 
     Exp: \citepP[P][]{STestSPICE3}, \citeP[P][]{SThe15essentialbuildingblocks} \\ \hline
    
    \multicolumn{4}{l}{  \footnotesize{\textsuperscript{$\ddagger$} The categories Eva, Exp, Opi, and Other have been introduced in Section 3.5.}}
    \end{tabular}
\end{table}

\begin{table}[h]
\centering
     \caption{ \label{tab:TestCodeQuality} Approaches to create and improve test script quality}
    \begin{tabular}{|p{2.2cm} | p{9.3cm}| p{1.8cm} | p{2.7cm} |}
    \hline
    Approach & Description &  AL citation\textsuperscript{$\ddagger$} &  GL citation\textsuperscript{$\ddagger$} \\ \hline

    Formal check & Check whether the given test scripts properly run on the SUT with respect to test cases and business requirements  & - & 
     Exp: \citepP[P][]{STestSPICE3} \\ \hline
    
    Use scripting techniques & E.g., structure scripting, shared scripts, data-driven testing, keyword-driven testing. Each technique has its drawbacks and strengths. Two or more techniques can be used together in practice.  &
     Exp: \citeP[P][]{SsoftwareTestAutomation} & 
     Exp: \citeP[P][]{SMaxim} \\ \hline 
    
    Statistic code analysis &  Analyze test codes to eliminate coding issues & 
     Exp: \citeP[P][]{SsoftwareTestAutomation}
    & 
     Exp: \citeP[P][]{SBasilios} \\ \hline 
    
    Development guidelines & Define development guidelines to train testers to comply with the same coding style, standards, and conventions &  - & 
     Exp: \citepP[P][]{SLessonsinTestAutomation}  \\ \hline
    
    Peer review & Do a peer review on test codes written by others to eliminate coding issues &  - & 
     Exp: \citepP[P][]{STestSPICE3}, \citeP[P][]{SLessonsinTestAutomation}\\ \hline 
    
    \multicolumn{4}{l}{  \footnotesize{\textsuperscript{$\ddagger$} The categories Eva, Exp, Opi, and Other have been introduced in Section 3.5.}}
   
    \end{tabular}
\end{table}

\paragraph{Arrange testware in good architecture.}
Arranging testware (e.g., test cases, test scripts, expected results, actual results, test logs, test data, and test reports) in good architecture was recommended by 14 sources (Table \ref{tab:factors}).  Many practitioners  \citeP[P][]{S7BestPracticesinTestAutomation}\citeP[P][]{SMeasurementofQualityoftheTesting}\citeP[P][]{SBrain}\citeP[P][]{SBasilios} reported that, based on their experience, the poor architecture makes it difficult to track, reuse, and maintain the growing testware in extensive test automation, and the associated costs would be high. Based on white papers  \citeP[P][]{SCentryLink}\citeP[P][]{SRanorex} from software organizations, some test tools have the default testware architecture that can be simply implemented. Recent experience studies \citeP[P][]{SMeasurementofQualityoftheTesting}\citeP[P][]{S7BestPracticesinTestAutomation} reported that, in a long-term perspective, it is better to design own testware architecture to fit an organization's own test automation context. Some practitioners learned from past experience and noted key issues to be considered in the design of testware architecture. Table~\ref{tab:TAarchitecture} summarized these key issues.

\begin{table}[ht]
\centering
     \caption{ \label{tab:TAarchitecture} Key issues to be considered in the design of testware architecture}
    \begin{tabular}{|p{2.5cm} | p{8.5cm}| p{2.3cm} |p{2.65cm} |}
    \hline
    Issue & Description &  AL citation\textsuperscript{$\ddagger$} & GL citation\textsuperscript{$\ddagger$} \\ \hline
    
    Access & How to let testers and test tools to easily locate and access required testware &  - &
    Exp: \citeP[P][]{SMeasurementofQualityoftheTesting}, \citeP[P][]{S7BestPracticesinTestAutomation}\\\hline 

    Scale & How to support the arrangement of testware that grow in size &  \textcolor{blue}{-} & 
     Exp: \citeP[P][]{SMeasurementofQualityoftheTesting}, \citeP[P][]{SsoftwareTestAutomation}\\ \hline
    
    Reuse &  How to maximize reuse of testware &  
    Exp: \citeP[P][]{Spersson2004establishment} &  -\\\hline 
    
    Multiple versions & How to manage the multiple versions of testware & 
    Exp: \citeP[P][]{Spersson2004establishment} &  - \\ \hline
    
    Differentiation & How to manage and distinguish the different types of testware & 
    Exp: \citeP[P][]{softwareTestAutomation} &  - \\ \hline
    
    Independence & If there is the need to test the SUT in multiple environments/hardware platforms, how to store and manage the independent testware of each environment/platform   & 
     Exp: \citeP[P][]{softwareTestAutomation} & 
     Exp: \citeP[P][]{SAhmed}
    \\ \hline
    
    \multicolumn{4}{l}{  \footnotesize{\textsuperscript{$\ddagger$} The categories Eva, Exp, Opi, and Other have been introduced in Section 3.5.}}
    \end{tabular}
\end{table}

\subsubsection{Test execution related practices.}

\paragraph{Prioritize automated tests for execution.} 9 sources suggested prioritizing the important automated tests for execution to enable early fault detection, fast feedback, and decreased costs for ensuring product quality. The literature presented two prioritization methods used in test automation practices, see Table~\ref{tab:TAPrioritization}. Based on cited sources  \citeP[P][]{SPractiTest}\citeP[P][]{STATCOE-Report}\citeP[P][]{SYang}\citeP[P][]{STestSPICE3}, Requirement based prioritization normally prioritize customer requirements at first according to some factors, e.g., customer demand, business value, fault impact of requirements. Based on cited sources \citeP[P][]{SsoftwareTestAutomation}\citeP[P][]{STestSPICE3}, Regression based prioritization is goal-oriented - It first sets the goals (e.g., early fault detection, product quality improvement, fewer costs) and then using test tools and algorithms to prioritize automated tests based on the relevant metrics. \citeP[P][]{STestSPICE3} presented case studies for both types of prioritization.

 \begin{table}[h]
\centering
     \caption{ \label{tab:TAPrioritization} Test prioritization}
    \begin{tabular}{|p{2.86cm} | p{8.7cm}| p{1.87cm} |p{2.65cm}|}
    \hline
    Type & Description & AL citation\textsuperscript{$\ddagger$} & GL citation\textsuperscript{$\ddagger$}\\ \hline
    Requirement based prioritization & Prioritize automated tests based on the priority of customer requirements. & - & 
    Exp: \citeP[P][]{SPractiTest}, \citeP[P][]{STATCOE-Report}, 
    \citeP[P][]{SYang}, 
    \citeP[P][]{STestSPICE3}  \\\hline 
    
    Regression based prioritization & Prioritize automated tests for regression testing, when the changes are made on the SUT/automated tests, after fixing the bugs, or finishing the updates of test environments.   &
    Exp: \citeP[P][]{SsoftwareTestAutomation} & 
    Exp: \citeP[P][]{STestSPICE3}\\\hline 
    
    \multicolumn{4}{l}{ \footnotesize{\textsuperscript{$\ddagger$} The categories Eva, Exp, Opi, and Other have been introduced in Section 3.5.}}
    
    \end{tabular}
\end{table}

\paragraph{Automate pre-processing and post-processing.} 
Two sources \citeP[P][]{SsoftwareTestAutomation} and \citeP[P][]{STATCOE-Report} suggested automating pre-processing (that sets prerequisites before the test execution) and post-processing (that deals with the aftermath of test execution) in test automation practices. They argued that pre-processing and post-processing related tasks are tedious and manually perform them consuming testing efforts. The advice on how to automate pre- and post- processing was identified in the study \citeP[P][]{SsoftwareTestAutomation}, an early source that proposed test automation guidelines based on authors' decades of test automation experience and field studies of case organizations. The study \citeP[P][]{SsoftwareTestAutomation} collected a set of pre-processing and post-processing related tasks that should be automated, see Table~\ref{tab:TAPrePro}. As advised by the study \citeP[P][]{SsoftwareTestAutomation}, these pre- and post- processing tasks can be implemented in shared scripts and command flies (e.g., command procedure, shell script, batch file) so that they can be automatically performed by test execution tools. The examples scripts and command files were given in the study \citeP[P][]{SsoftwareTestAutomation}.

\begin{table}[h]
\centering
     \caption{\label{tab:TAPrePro} Pre-processing and post-processing tasks [P7]}
    \begin{tabular}{|p{2.4cm} | p{10cm} |}
    \hline
     & Tasks  \\ \hline
    Pre-processing &  Create files, test data, databases; Prepare test environments; Format test data and test scripts; Install libraries and dependencies.   \\\hline 
    
    Post-processing &  Delete files, databases, datasets; Reorganize test results; Format outcomes.  \\\hline 
    
    \end{tabular}
\end{table}

\subsubsection{Verdicts related practices.}
\label{sec:results_verdicts}

\paragraph{Automate test oracles.} 
Two sources  \citepP[P][]{STATCOE-Report}\citeP[P][]{Sgarousi2016} studied the effect of test oracle on test automation maturity. The source  \citeP[P][]{Sgarousi2016} is a SLR. The source \citeP[P][]{STATCOE-Report} is a technical report from STAT COE center that supports U.S. government testing and evaluation programs. Based on these two sources, test oracle is a mechanism to verify the correctness of outputs of automated test cases concerning expect results. Both sources suggested that test oracle automation can reduce human efforts in verifying the outputs from executing test cases and thus increase the testing scale in a short test cycle. Yet, these two sources did not include the advice on how to conduct test oracle automation in practice.

\paragraph{Analyze test automation results efficiently and effectively.} 
11 sources (Table \ref{tab:factors}) noted that efficiently and effectively analyze test automation results is important to mature test automation. As reported by practitioners  \citeP[P][]{S7BestPracticesinTestAutomation}\citeP[P][]{Sskip}, in their past practices, when the time spent on analyzing test automation results exceeded the time saved from executing automated tests, test automation slowed the development work and it was not worth the cost. Experience studies \citeP[P][]{S7BestPracticesinTestAutomation}\citeP[P][]{SRanorex} reported that ineffective test automation results harm software quality. Many sources (include experience \& opinion studies) advised the approaches to efficiently and effectively analyze test automation results, as summarized in Table \ref{tab:TAresultsAnalysis}. The study \citeP[P][]{SsoftwareTestAutomation} presented examples to implement the suggested approaches (Notifications, Big picture, and Keep history) on test execution tools.

\begin{table}[h]
\centering
     \caption{ \label{tab:TAresultsAnalysis} Approaches to analyze test automation results}
    \begin{tabular}{|p{3.2cm} | p{7.8cm}| p{2.4cm} | p{2.65cm} |}
    \hline
     Approaches & Description & AL citation\textsuperscript{$\ddagger$} & GL citation\textsuperscript{$\ddagger$} \\ \hline
     Interpret and classify test automation results &  Analyze failed tests to find the root reason for failure and classify the results to prevent potential incidents. & -&
     Exp: \citepP[P][]{STestSPICE3}, \citeP[P][]{SThe15essentialbuildingblocks}, \citeP[P][]{STATCOE-Report}\\\hline 
    
    More than ``pass" or ``fail" &  Store and review the artifacts (logs, screenshots, comparisons, or video recordings of test runs, and others generated from executing automated tests) to get complement information for debugging and fixing issues. & 
    Exp: \citeP[P][]{Spersson2004establishment}& 
    Exp: \citeP[P][]{STestSPICE3}, \citeP[P][]{SUdi}, \citeP[P][]{STATCOE-Report} \\\hline 
    
    Notifications & Set the notifications (on test execution tools) to alarm the failures of critical automated tests, so that the priority can be given to analyze and solve the failures of critical automated tests when receiving the notifications.& 
    Exp: \citeP[P][]{SsoftwareTestAutomation}& 
    Exp: \citepP[P][]{Sskip}\\\hline
    
    Tool support & Use test tools that can give a clear overview of each step of the test flow so that failures can be quickly identified. & 
    - &
    Exp: \citeP[P][]{STestinginaDevOpsMindset}\\\hline
    
    Smoke tests & Run smoke tests on automated test suites incrementally to expose reasons for failures. & - & Opi: \citeP[P][]{SScott}\\\hline
    
    Big picture &  In addition to analyze a single test run results, it is essential to combine test automation results collected from different sources (e.g., across multiple test tools, test runs, configurations, integration builds, and milestones) into a big picture view of outcomes.& Exp:\citepP[P][]{Sstbox3}, \citeP[P][]{SsoftwareTestAutomation}  & - \\\hline

    Keep history & Store test automation results for a period of time to enable progress tracking, regression identification, and flaky tests identification. & Exp: \citepP[P][]{STATCOE-Report}, \citeP[P][]{SsoftwareTestAutomation}& - \\\hline
    
    \multicolumn{4}{l}{  \footnotesize{\textsuperscript{$\ddagger$} The categories Eva, Exp, Opi, and Other have been introduced in Section 3.5.}}
    
    \end{tabular}
\end{table}

\paragraph{Report useful test automation results to key stakeholders.} 
As mentioned in
6 sources (Table \ref{tab:factors}), once test runs were executed, the next thing is test automation result reporting, which is the approach to make test automation results visible to stakeholders and inform them about their work. Based on test maturity models TestSPICE 3.0 \citepP[P][]{STestSPICE3} and STBox 3.0 \citepP[P][]{Sstbox3}, effective reporting can enable the track of test progress, make the tests reliable, and increase the quality of automated tests. They advised organizations to automatically report test automation results to key stakeholders using test tools. However, practitioners \citepP[P][]{SsoftwareTestAutomation} claimed that, based on their decades' test automation experience, not all reports generated by test tools are useful for all stakeholders, and thus, it is important to have a selection and only send the required reports to the right stakeholders. The study  \citepP[P][]{SsoftwareTestAutomation} illustrated examples of what types of reports different stakeholders would prefer: decision makers might be interested in quality reports and progress reports to observe how test automation supports software development; Developers would like to get bug reports, crash reports, and error reports in order to get feedback about their development work and identify causes and fixes; Testers would also be interested in crash reports and error reports to monitor the test automation status and performance.

\subsubsection{Measurement related practices}
\label{sec:resultsMeasrument}
 \paragraph{Use the right test automation metrics.} 
10 sources (Table \ref{tab:factors}) recommended using the right test automation metrics to quantitatively measure test automation performance and maturity improvement actions. Based on many experience studies \citeP[P][]{SsoftwareTestAutomation}\citeP[P][]{STATCOE-Report}\citeP[P][]{SPlanningforLongtermTestAutomationSucess}\citeP[P][]{SSealights}\citeP[P][]{SPlanningforLongtermTestAutomationSucess}, the right test automation metrics refers to the ones ``that are objective, measurable, and meaningful to serve the test automation goals within an organization''; the wrong metrics might waste efforts and obtain irrelevant/misleading data to measure test automation performance and maturity improvement actions. Several sources (Table \ref{tab:TAMetrics}) presented a collection of example test automation metrics. The study \citeP[P][]{SsoftwareTestAutomation} observed each of these test automation metrics in several cases of software organizations and presented the details on the usage of each metric in practice. Based on the experience of several practitioners  \citeP[P][]{SPlanningforLongtermTestAutomationSucess}\citeP[P][]{SSealights},  some tools (e.g., SonarQube, GitLab) also provide the collection of test automation metrics and metric data can be collected and recorded automatically. Experience studies  \citeP[P][]{SsoftwareTestAutomation}\citeP[P][]{SSealights}\citeP[P][]{STestinginaDevOpsMindset} suggested organizations define and customize test automation metrics based on their own needs, but how to do that was not introduced in these studies.

\begin{table}
     \caption{ \label{tab:TAMetrics} Test automation metrics}
    \begin{tabular}{ |p{2cm}| p{9.7cm}| p{1.9cm}|p{1.9cm}|}
    \hline
    \textbf{Measurement} & \textbf{Example metrics} & AL citation\textsuperscript{$\ddagger$} & GL citation\textsuperscript{$\ddagger$}  \\\hline
     
    Test automation coverage
     & Lines of code tested by test automation; \newline
      Increased coverage of expected operational paths and use cases;     \newline
      Percentage of test coverage achieved by automated tests. & 
      Exp: \citeP[P][]{SsoftwareTestAutomation} &
      Exp: \citeP[P][]{SSealights} \newline
      Opi: \citeP[P][]{SMatthews}
      \\ \hline
     
  Test \newline efficiency 
     & Average time to design and execute automated tests; \newline
      Average time to run test automation sequence; \newline
     Average time to provide feedback to development works; \newline
     Percentage of automated tests passed, skipped, and failed w.r.t. total number of automated tests planned to run. & 
     Exp: \citeP[P][]{SsoftwareTestAutomation} &
     Exp: \citeP[P][]{SSealights} \newline 
     Opi: \citeP[P][]{SMatthews}
     \\ \hline
     
  Test \newline effectiveness &
      Percentage of useful test automation results; \newline
      Defect detection percentage = defects found by automated tests / total known defects; \newline
      \quad Defect fix percentage = defects fixed before release / all defects found. & 
     Exp: \citeP[P][]{SsoftwareTestAutomation} &
     Exp: \citeP[P][]{SSealights} \\\hline
      
 Product \newline quality &
     Number of defects in production; \newline
     \quad Satisfaction rate of users & Exp: \citeP[P][]{SsoftwareTestAutomation} &
     Exp: \citeP[P][]{SSealights}
     \\ \hline
     
  Maintenance efforts &
     Number of defects in production; \newline
     Reuse rate of test automation artifacts; \newline
     The average elapsed time to update automated tests. & 
     Exp: \citeP[P][]{SsoftwareTestAutomation} &
     Exp: \citeP[P][]{SSealights}
     \\\hline
     
    \multicolumn{4}{l}{  \footnotesize{\textsuperscript{$\ddagger$} The categories Eva, Exp, Opi, and Other have been introduced in Section 3.5.}}
     
    \end{tabular}
\end{table}

\subsubsection{SUT related practices.} 
\label{sec:SUT_practice}

\paragraph{Design the SUT for automated testability.} 12 sources (Table \ref{tab:factors}) mentioned that it is important to design the SUT for automated testability. An academic case study \citeP[P][]{Sborjesson2012} on test automation with visual GUI tools found that the SUT testability affects test automation maturity. As reported by several experience studies ~ \citepP[P][]{S7BestPracticesinTestAutomation}\citeP[P][]{SPractiTest}\citeP[P][]{SBrain}, many SUTs might not be testable for test automation, e.g., the SUT that changes often, is not stable, is difficult to write test scripts on that, or has low availability and running speed. As concluded in a SLR \citepP[P][]{Swiklund2017impediments} on the impediments of test automation maturity, the SUT with low automated testability might need an extensive workload to develop test automation, test automation functions may fail on some parts of the SUT, and thus software development productivity suffers. However, from cited sources in Table \ref{tab:factors}, we did not identify the advice on how to design the SUT for automated testability.

\subsubsection{Technology related practices.} 
\label{sec:technology_practice}

\paragraph{Adopt new technologies.} 
4 sources (Table \ref{tab:factors}) suggested adopting new technologies in test automation practices. Based on World Quality Reports \citeP[P][]{SWQR17}\citeP[P][]{SWQR19}\citeP[P][]{SWQR20} that annually surveyed global quality and testing practices and analyzed the trends, in the area of test automation, there was the evolution when new technologies were commoditized in the industry: in the early years, test automation was only used to do simple regression tests with reply and record tools; These days, test automation technologies have become more complex and that are being used to carry out different levels of tests (unit test, integration test, performance test) and support diverse testing activities (e.g., test design, test execution, measurements, test environment provision, defect prediction, test management); In the future, advanced technologies have the promise to build more intelligent test automation that is more effective and efficient to test more. As reported by industrial experts in these reports, based on their estimation, software organizations staying with old technologies may lose the competencies in the future; Thus, they suggested software organizations follow test automation technical trends in the industry and watch for opportunities for action. However, from cited sources (Table \ref{tab:factors}) of this best practice, the advice on how to adopt new technologies within an organization was not identified.

\label{sec:discussion}
 \section{Discussions}\label{sec:dicussion_sec}
We summarize and discuss the study finding in Section \ref{sec:SumaaryStudyResults}, explore the implications to research in Section~\ref{sec:FutureResearch} and the implications to practice in \ref{sec:implicationsPractice}, and examine threats to validity in Section \ref{sec:threats}.

\subsection{Summary and discussion of study findings}
\label{sec:SumaaryStudyResults}
This MLR has the objective to survey and synthesize the guidelines given in the literature for test automation maturity improvement. To solve the objective, from a large pool of sources, we selected and reviewed 81 primary studies (including 55 GL and 26 AL sources) on this topic. Many of our primary studies are experience reports (n=34) that proposed test automation heuristics \& guidelines (Figure~\ref{fig:bubbleplot}). Next, we answer our research questions and discuss study findings on them.

\paragraph{RQ1. Which test automation best practices are given in the literature? } \textit{As an answer to RQ1, from the literature, we extracted 26 test automation best practices and grouped them into 13 key areas, as presented in Table \ref{tab:factors}}. As described in Section \ref{sec:resultsPractices}, there are only 6 best practices whose positive effect on maturity improvement have been validated by academic evaluation studies using formal empirical methods. Future academic evaluation studies are needed to validate the effect of other best practices on test automation maturity improvement. 

Our previous work \cite{Wang2019} is the only prior academic study that also proposed test automation best practices. Our previous work collected test automation best practices from 18 test maturity models, which are used to guide test automation practices in the industry. Six technical related best practices proposed in this MLR were not presented in our previous work: Set up good test environments, Create high-quality test data, Develop high-quality test scripts, Automate test oracles, Analyze test automation results efficiently and effectively, Adopt new technologies. This denotes that, the recent AL and GL (reviewed in this MLR) proposed test automation best practices that were not covered in test maturity models (reviewed in our previous work). Once the positive effect of these six technical related best practices on test automation maturity has been validated with cross-site empirical evidence, there is a need to update test maturity models that miss these best practices. Additionally, one management related best practice "Establishing a test organization (like a test team or a department) to assemble people to perform test automation tasks" was presented in some test maturity models (reviewed in our previous work) but was not identified from the current AL and GL (reviewed in this MLR). The possible reason could be the wide adoption of continuous integration drives the change. In continuous integration contexts, development and test automation are integrated, and thus, in many cases, developers perform both development and test automation tasks \cite{garousi2020}. Yet, further studies are needed to explore the details.

As discussed in Table \ref{tab:relatedSecondary} (Section \ref{sec:relatedwork}), to provide the guidelines for test automation maturity improvement, Rodrigues et al.~\cite{rodrigues2016} proposed a taxonomy of test automation success factors and Wiklund et al.~\cite{wiklund2017impediments} proposed a taxonomy of maturity impediments. We found that, our taxonomy of test automation best practices in this MLR and their taxonomy of test automation success factors and maturity impediments can be linked together. Table~\ref{tab:LinkToRodrigues} shows which key areas of our best practices can be linked to each success factor in Rodrigues et al.'s study. Table~\ref{tab:LinktoKristian} shows which key areas of our best practices can be linked to each category of maturity impediments in Wiklund et al.' study. Compared to their work, our MLR covers more test automation key areas. Rodrigues et al.'s study did not present success factors in Knowledge transfer, Test environment, Test automation requirements, Test execution, Verdicts, and Technology key areas, see Table~\ref{tab:LinkToRodrigues}. Wiklund et al.'s study did not identify maturity impediments in Test execution, Measurement, and Technology key areas, see Table~\ref{tab:LinktoKristian}. Additionally, referring to the gap analysis in Table \ref{tab:relatedSecondary}, their studies did not present how to address success factors and impediments (they proposed) in practice. Our MLR complements their work.  Our best practices with the collected advice (in RQ2) on how to conduct them can be used to address relevant success factors and maturity impediments in their work.

\begin{table}[htpb]
\caption{ \label{tab:LinkToRodrigues} Rodrigues et al.'s \cite{rodrigues2016} success factors v.s. Our key areas of best practices}

    \begin{tabular}{ |p{2.5cm} | p{0.7cm}  |p{0.7cm} | p{0.7cm} |p{0.7cm} | p{0.7cm} | p{0.7cm} | p{0.7cm} |p{0.7cm} | p{0.7cm}| p{0.7cm} | p{0.7cm}| p{0.7cm} | p{0.7cm}|}
    \hline
    
     & \rotatebox{90}{\parbox{2.9cm}{Test Automation \\ strategy}} & \rotatebox{90}{\parbox{2.9cm}{Resources }} & 
     \rotatebox{90}{\parbox{2.9cm}{Test \\ organization }} & 
     \rotatebox{90}{\parbox{2.9cm}{Knowledge \\transfer }} & 
      \rotatebox{90}{\parbox{2.9cm}{Test tools }} & 
     \rotatebox{90}{\parbox{2.9cm}{Test \newline environment }} & 
     \rotatebox{90}{\parbox{2.9cm}{Test  automation \\ requirements }} & 
     \rotatebox{90}{\parbox{2.9cm}{Test design }} &
     \rotatebox{90}{\parbox{2.9cm}{Test execution }} & \rotatebox{90}{\parbox{2.9cm}{Verdicts}} & 
     \rotatebox{90}{\parbox{2.9cm}{Measurements}} & 
     \rotatebox{90}{\parbox{2.9cm}{SUT}} & 
     \rotatebox{90}{\parbox{2.9cm}{Technology}} \\\hline
    
    Feasibility assessment &  \cellcolor{gray!25} Yes &  &    &  &   & & & & & & & &   \\ \hline
    
    Testability Level of the SUT &  &  &    &  &   & & & & & & &  \cellcolor{gray!25} Yes &  \\ \hline
    
    Resource availability &  & \cellcolor{gray!25} Yes &    &  &   & & & & & & & &   \\ \hline
    
    Resource Reusability &  & \cellcolor{gray!25} Yes &    &  &   & & & & & & & &   \\ \hline
    
    Well Defined Test Process &  \cellcolor{gray!25} Yes &  &    &  &   & & & & & & & &   \\ \hline

    Scalability &   &  &    &  &   & & &  \cellcolor{gray!25} Yes &  & & & &    \\ \hline
    
    Maintainability &   &  &    &  &   & & & \cellcolor{gray!25} Yes &  & & & &  \\\hline
    
    Automation Tool Acquisition&   &  &    &  &  \cellcolor{gray!25} Yes &  & &  &  & & & &  \\ \hline
    
    Quality Control &   &  &    &  &   & & &   & & & \cellcolor{gray!25} Yes &  & \\ \hline
    
    Dedicated and Skilled Team &   &  & \cellcolor{gray!25} Yes  &  &   & & &   & & &  &  &  \\ \hline
    
    Automation Planning/Strategy & \cellcolor{gray!25} Yes  &  & \cellcolor{gray!25} Yes  &  &   & & &   & & &  &  &   \\ \hline

\end{tabular}
\vspace{-3mm}
\end{table}

\begin{table}[ht]
\caption{\label{tab:LinktoKristian} Wiklund et al.'s \cite{wiklund2017impediments} categories of maturity impediments v.s. Our key areas of best practices}
    \begin{tabular}{ |p{2.4cm} | p{0.7cm}  |p{0.7cm} | p{0.7cm} |p{0.7cm} | p{0.7cm} | p{0.7cm} | p{0.7cm} |p{0.7cm} | p{0.7cm}| p{0.7cm} | p{0.7cm}| p{0.7cm} | p{0.7cm}|}
    \hline
    
    & \rotatebox{90}{\parbox{2.9cm}{Test Automation \\ strategy}} & \rotatebox{90}{\parbox{2.9cm}{Resources }} & 
     \rotatebox{90}{\parbox{2.9cm}{Test \\ organization }} & 
     \rotatebox{90}{\parbox{2.9cm}{Knowledge \\transfer }} & 
      \rotatebox{90}{\parbox{2.9cm}{Test tools }} & 
     \rotatebox{90}{\parbox{2.9cm}{Test \newline environment }} & 
     \rotatebox{90}{\parbox{2.9cm}{Test  automation \\ requirements }} & 
     \rotatebox{90}{\parbox{2.9cm}{Test design }} &
     \rotatebox{90}{\parbox{2.9cm}{Test execution }} & \rotatebox{90}{\parbox{2.9cm}{Verdicts}} & 
     \rotatebox{90}{\parbox{2.9cm}{Measurements}} & 
     \rotatebox{90}{\parbox{2.9cm}{SUT}} & 
     \rotatebox{90}{\parbox{2.9cm}{Technology}} \\\hline
    
    Behavioural effects &  \cellcolor{gray!25} Yes &  &  \cellcolor{gray!25} Yes  &  &   & & & & & & & &   \\ \hline
    
    Business and planning & \cellcolor{gray!25} Yes & \cellcolor{gray!25} Yes & \cellcolor{gray!25} Yes   &  &   & & & & & & &  &  \\ \hline
    
    Skills &  &  &  \cellcolor{gray!25} Yes & \cellcolor{gray!25} Yes &   & & & & & & & &   \\ \hline
    
    Test system &  &  &    &  &  \cellcolor{gray!25} Yes & \cellcolor{gray!25} Yes & \cellcolor{gray!25} Yes & \cellcolor{gray!25} Yes & & \cellcolor{gray!25} Yes & & &   \\ \hline
    
    SUT &  &  &    &  &   & & & & & & & \cellcolor{gray!25} Yes &   \\ \hline

\end{tabular}
\vspace{-3mm}
\end{table}

\paragraph{RQ2. What advice was given in the literature about how to conduct proposed test automation best practices? } \textit{From the literature, we identified many pieces of advice (in forms of implementation or improvement approaches, actions, technical techniques, concepts, experience-based heuristics) on how to conduct proposed test automation best practices in RQ1.} We have several observations as described below:

 We can see from Section \ref{sec:resutls_RQ2}, most pieces of advice on how to conduct proposed best practices were identified from experience studies and their effectiveness needs to be evaluated with cross-site empirical evidence using formal empirical methods. Only some pieces of advice on how to "Select the right test tools" (Section~\ref{sec:testtools}) have been evaluated by academic studies with cross-site empirical evidence: to support organizations to select right test tools, the literature has defined the concept of "the right test tools" and proposed selection criteria, and the content validity of the concept and selection criteria have been evaluated by academic studies that surveyed cross-site test professionals.
 
Several pieces of advice on how to conduct some best practices are conflicting. Qualitative studies are needed to explain the conflicts. First, as the advice for "Involving key stakeholders in strategy development" (Section \ref{sec:resultStrategyDefine}), some test maturity models (proposed by authorized organizations) suggested the formal approach for general software development context - managers lead in working with key stakeholders to formally discuss the main topics on a test automation strategy. Meanwhile, some practitioners viewed that, based on their experience, in agile contexts, it can be done in an informal approach - the main topics on a test automation strategy can be aware of at any time and informal discussion among key stakeholders will occur when necessary. Second, as the advice for "Select the right test tools" (Section \ref{sec:testtools}), many scholars suggested selecting test tools against pre-defined selection criteria, while some practitioners viewed that - based on their experience - selecting test tools against pre-defined criteria is less useful than selecting each test tool with an experimentation mindset. Third, as the advice for "Define test automation requirements" (Section \ref{sec:RQ2_results_requirements}), some test maturity models suggested that both test automation requirements and manual testing requirements are software testing requirements, while recent industry reports mentioned that test automation requirements are different from manual testing requirements.

The pieces of advice on how to conduct some best practices still need further qualitative analysis. First, as the advice for "Share available test automation knowledge''(Section \ref{sec:knowledgeTransfer}), prior studies discussed different types of shareable test automation knowledge and advised places to share it, but did not include details on how to share these types of shareable test automation knowledge through advised places. Second, as the advice for "Define test automation requirements" (Section \ref{sec:RQ2_results_requirements}), recent industry reports mentioned that many organizations are using machine learning techniques to automatically extract functional requirements from existing test cases, but details on how they do that were not introduced. Third, as the advice for "Use the right test automation metrics" (Section \ref{sec:resultsMeasrument}), the literature defined the concept of "right test automation metrics" and illustrated a collection of example ones, but it did not mention how to define and customize test automation metrics based on own needs of an organization.

We did not identify the advice on how to conduct some management related best practices: ``Acquire enough management support for test automation'' (Section \ref{sec:testOrganizationFactors}), ``Keep test professionals motivated'' (Section \ref{sec:testOrganizationFactors}), ``Promote collaboration'' (Section \ref{sec:testOrganizationFactors}), ``Allow time for training and learning curve'' (Section \ref{sec:knowledgeTransfer}), ``Adopt new technology'' (Section \ref{sec:technology_practice}). However, we found that SE research has given advice for adopting similar management related best practices in SE contexts. For example, researchers \cite{ragu2004} have advised the main steps to acquire management support for SE activities within an organization: Engage key managers, Express expected results, Frame a project in the context of organization objectives, Gather feedback, and Share good news. The study \cite{whitehead2007} proposed a process to promote collaboration in SE, from bringing awareness of collaboration in the organizational culture to establishing real-time collaboration with tools, environments, and infrastructure to facilitate informal and formal communications. The study \cite{verner2014} observed that SE practitioners can be motivated by several motivators: Increased salary, Career promotion, Technical growth, Personal identity with tasks, Autonomy, Working environment. Software process model CMMi \cite{team2002capability} proposed the main activities of an organizational training program: Identify training needs, Providing training, Establishing and maintaining the training capability and records, Assessing training effectiveness. Nemoto et al. \cite{nemoto2010} described four variables that should be considered in order to successfully adopt the new computer technology within an organization: Attributes of innovation, External environment, Suppliers, Organizational characteristics. As test automation is a component of SE, with some modifications, such pieces of advice from SE research can be immigrated into the test automation context, although evaluation and guidance studies are required to validate the effect.

The advice on how to conduct technical related best practices "Design the SUT for automated testability" (Section \ref{sec:technology_practice}) and "Automate test oracles" (Section \ref{sec:results_verdicts}) was not found from our primary studies. However, to our best knowledge, there exist technical related AL on these two practices. To extend the study, we applied the search string "SUT testability automat*" and ``test oracle automat*'' in Google Scholar searching for technical AL on these two practices. We reviewed relevant studies and found that much work remains to be done in order to support the industry to conduct these two practices. First, we observed that, the approaches to design automated testability are very diverse and highly dependent on the application domain of the SUT and test types. Some technical approaches have been proposed to design the SUT for automated testability for embedded systems, GUI test automation, objective-oriented systems \cite{mariani2014,kanstren2008,kansomkeat2008analysis}. Yet, technical approaches to design automated testability of SUT for other application domains (e.g., mobile applications, software-defined systems) and test types (e.g. unit testing, integration testing, performance testing) are still needed \cite{mariani2014,kanstren2008,yoo2012regression,kansomkeat2008analysis}. Second, we observed that technical techniques have been proposed to support test oracle automation. Barr et al.~\cite{harman2013} has constructed a repository of 694 publications on test
oracles and studied the existing technical techniques for test oracle automation. They found that the existing technical techniques (including modeling, specifications, contract-driven development, and metamorphic testing) are not sufficient, as the human intervention still need to provide the final source of test oracle information, e.g., informal specifications, expectations, norms, and domain-specific information - as such, test oracle automation still needs relatively high costs, and thus, how to keep the benefits while reducing the costs is still a challenge. This calls for more technical studies on test oracle automation.

\subsection{Implications to research} 
\label{sec:FutureResearch}

This MLR narrows the gap between practice and research - the industry needs to improve test automation maturity improvement, while the research effort on surveying and synthesizing the guidelines given in the literature for improving test automation maturity is limited. This MLR spots research challenges and opportunities around four topics:

\paragraph{Empirical studies.} More empirical studies are acutely needed. In this MLR, we only identified 10 evaluation studies presenting empirical results, despite the industry relevance of this research topic. This MLR points out empirical study opportunities. First, there are only six test automation best practices whose positive effect on maturity improvement have been evaluated by academic studies using formal empirical methods, and thus, future academic studies are needed to evaluate the effect of other best practices on maturity improvement as well. Researchers can adopt such best practices in a given industrial context, e.g., in a specific domain like embedded software systems, in small and medium size organizations, or in a continuous integration context, and then observe the effect. The relationships between different best practices and the intended effects also can be observed. Second, in this MLR, we collected many pieces of advice (from different sources) on how to conduct the proposed best practices, while most of them come from experience studies. Studies are needed to observe the consequence of following the collected pieces of advice to conduct the proposed best practices in different organizations and as such evaluate the effectiveness of these pieces of advice.

\paragraph{Qualitative studies.} 26 test automation best practices in this MLR notice the important research topics for test automation research. Due to the scope of this MLR, we only provided an overview of the pieces of advice (proposed by prior researchers and practitioners) on how to conduct these best practices. Thus, we suggest more case studies to observe how these 26 best practices can be conducted against the collected pieces of advice. For example, case studies can be conducted on the adoption of the suggested approaches to set up/ improve test environments (Table \ref{tab:TATestEnvrionmentAppraoch}), create and improve test script quality (Table \ref{tab:TestCodeQuality}), and analyze test automation results (Table \ref{tab:TAresultsAnalysis}) within different organizations. Additionally, referring to the observations on study results of RQ2 in Section~\ref{sec:SumaaryStudyResults}, future studies are needed to explain the conflicts among the pieces of advice on how to conduct certain best practices (``Involving key stakeholders in strategy development'', ``Select the right test tools'', and ``Define test automation requirements''), and add further qualitative analysis to the pieces of advice on how to conduct certain best practices (``Share available test automation knowledge'', ``Define test automation requirements'', ``Use the right test automation metric'').

\paragraph{Technical studies.} Referring to observations to study results of RQ2 (Section \ref{sec:SumaaryStudyResults}), technical studies are required to (1) develop approaches for designing automated testability of the SUT in different application domains and test types, (2) and advance test oracle automation techniques with least human efforts. Besides, as test automation is technology-driven and technology maturity is an important dimension of test automation maturity (see definition in Section~\ref{sec:conceptOfMaturity}), more technical studies are required to advance the next level of test automation maturity.

\paragraph{Maturity models.} Test maturity models are being used to guide test automation practices in the industry. However, referring to observations to study results of RQ1 (Section \ref{sec:SumaaryStudyResults}), this MLR identified six technical related best practices that were not found in test maturity models (reviewed in our previous work \cite{Wang2019}): Set up good test environments, Create high-quality test data, Develop high-quality test scripts, Automate test oracles, Analyze test automation results efficiently and effectively, Adopt novel technologies. Once the positive effect of these six technical related best practices on improving test automation maturity has been evaluated, future work is needed to update test maturity models that miss these best practices. Additionally, many researchers denoted the need to develop test automation maturity models, since test maturity models focus more on manual testing than test automation~\cite{wiklund2015Dhd,eldh2014towards,furtado2014}. Other researchers can use our study results of this MLR as a base to develop test automation maturity models. The application of the updated models or new models should be evaluated by research and practice.

\subsection{Implications to practice}
\label{sec:implicationsPractice}
Our MLR proposed 26 test automation best practices and collected the pieces of advice on how to conduct them. It has implications for organizations that are doing test automation or improving test automation maturity. To be more specific, by consulting our study results, organizations may find 26 test automation best practices to follow during the planning and the implementation of test automation. Additionally, they can also compare their existing test automation practices with these 26 best practices to assess the current state of their test automation practices. Our MLR identified the pieces of advice (from prior scholars and practitioners) on how to conduct the proposed 26 best practices. Referring to such pieces of advice around these best practices, organizations may get hints about the next steps for improving their test automation maturity. 

\subsection{Threats to validity}
\label{sec:threats}
 We identified the main threats that may affect the validity of this MLR according to the guidelines from Petersen and Gencel~\cite{petersen2013}. Their guidelines provide a taxonomy of validity threats in SE research. 

\paragraph{Descriptive validity.} Descriptive validity determines ``the extent to which the observations of a study are described objectively and accurately \cite{petersen2013}''. In SLR studies, researchers ignore important data on primary studies may lead to a threat to descriptive validity. To control this threat, we defined review questions and specified what data should be extracted from primary studies. Accordingly, we coded all relevant findings from primary studies. During the coding process, the original sources were always reviewed to ensure the validity of codes.

\paragraph{Theoretical validity.}
Theoretical validity refers to the extent to which a study reflects what it intends to reflect \cite{petersen2013}. Some relevant studies were excluded. This may threaten the theoretical validity of this MLR. We restricted our search to English language sources, though some relevant studies have been published in other languages. Additionally, human decisions on search selection may imperil the theoretical validity. To control that, we ensured the source selection process involves two authors. Sources were finally selected according to the selection results of two authors, and final decisions were shown to other co-authors for the review. 

\paragraph{Interpretive validity.}
Interpretive validity focus on how conclusions are reasonably drawn on the given data in a study~\cite{petersen2013}. Researcher bias may introduce a threat to this type of validity. The interpretation of data may depend on the experience and thoughts of researchers of this paper. Though the interpretation of data was made by the first author, the conclusions were reviewed by other authors. The second and fourth authors are experienced in conducting MLRs and have published extensively in the field of software
testing or SE. Besides, when presenting conclusions, examples were used in many places to provide more information.

\paragraph{Generalizability.}
Generalizability concerns the extent to which the results of a study can be generalized~\cite{petersen2013}. In this MLR, we take care of how study results can be generalized to the industrial and academic contexts. Practitioners need to consider individual differences in their test automation practices when consulting study results of this MLR. Besides, SE is a rapidly evolving discipline and the generalizability of study results depends on the current context \cite{petersen2013}. Hence, the threat lies in whether our research findings can be generalized in the field of SE after a couple of years later.

\paragraph{Repeatability.}
Repeatability refers to the extent to which a study is repeatable~\cite{petersen2013}. To avoid threats to repeatability, we carried out this MLR by following `the guidelines for including grey literature  and conducting multivocal literature reviews' from Garousi \textit{et al.}~\cite{garousi2018guidelines}. The search strategy was defined and it was accordingly followed to conduct this MLR. Intermediary results of research stages were stored in spreadsheets and NVivo files. The whole research process was reported in detail.

\section{Conclusion}
\label{sec:conclusion}
Given the popularity of test automation and the large investments that might be wasted with negative outcomes from being immature, software organizations must improve test automation maturity. This MLR set the study objective to survey and synthesize the guidelines given in the current literature for test automation maturity improvement. In solving the study objective, this MLR selected and reviewed 81 primary studies consisting of 26 AL and 55 GL sources on this topic. From primary studies, this MLR extracted 26 test automation best practices and collected many pieces of advice (in forms of implementation/improvement approaches, actions, technical techniques, concepts, experience-based opinions) on how to conduct these best practices.

This MLR has four main contributions. First, it proposed 26 test automation best practices to suggest steps for improving test automation maturity. Practitioners can consult these best practices to improve test automation maturity or build test automation from scratch, see details in Section \ref{sec:implicationsPractice}. Second, it narrows the gap between practice and research - the industry needs to improve test automation maturity, while the current research lacks the synthesis on this topic. Third, this MLR provides a centralized knowledge base of existing guidelines given by prior scholars and practitioners for test automation maturity improvement. This centralized knowledge base can be used to get an overview of research advancements in this field, and frost future research topics that merit attention. Last, this MLR identified research challenge and opportunities around four topics: empirical studies to validate the study results of this MLR, qualitative studies to explain some details and conflicts, technical studies to advance test automation technology, and maturity model related studies, see Section \ref{sec:FutureResearch} for details.

\section*{Acknowledgements} 
This work is supported by TESTOMAT Project (ITEA3 ID number 16032) funded by Business Finland under Grant Decision ID 3192/31/2017, and the foundation of Tauno Tönning (project ID 20210086). 

\bibliographystyle{wileyj}
\bibliography{main}

\section*{Appendix A. Primary studies} 
\bibliographystyleP{wileyj}
\renewcommand{\bibsection}{\textbf{P1-P26 are academic literature sources. P27-P81 are grey literature sources.} }
\bibliographyP{refs-etc}

\end{document}